\documentclass{aa}  
\usepackage{graphicx}
\usepackage{txfonts}
\newcommand{\lsim}{\ \raise -2.truept\hbox{\rlap{\hbox{$\sim$}}\raise
    5.truept\hbox{$<$}\ }} \newcommand{\gsim}{\ \raise
  -2.truept\hbox{\rlap{\hbox{$\sim$}}\raise 5.truept\hbox{$>$}\ }}
\newcommand{\nodata}{...}
\newcommand{\gi}{$(g{-}i)$\ }
\newcommand{\gr}{$(g{-}r)$\ }

\newcommand{\feh}{$[Fe/H]$\ }
\newcommand{\ur}{$(u{-}r)$\ }

\def\Fig{\mbox{Figure~}}

\begin{document} 
   \title{The Fornax Deep Survey with VST. IX}

   \subtitle{The catalog of sources in the FDS area, with an example
     study for globular clusters and background galaxies}


   \author{Michele Cantiello\inst{1}\and
Aku Venhola\inst{2}\and
Aniello Grado\inst{3}\and
Maurizio Paolillo\inst{4,5}\and
Raffaele D'Abrusco\inst{6}\and
Gabriella Raimondo\inst{1}\and
Massimo Quintini\inst{1}\and
Michael Hilker\inst{7}\and
Steffen Mieske\inst{8}\and
Crescenzo Tortora\inst{9}\and
Marilena Spavone\inst{3}\and
Massimo Capaccioli\inst{4}\and
Enrica Iodice\inst{3,7}\and
Reynier Peletier\inst{10}\and
Jes\'us Falc\'on Barroso\inst{11,12}\and
Luca Limatola\and
Nicola Napolitano\inst{13,3}\and
Pietro Schipani\inst{3}
Glenn van de Ven\inst{14}\and
Fabrizio Gentile\inst{4}\and
Giovanni Covone\inst{4,3,5}
}

 \institute{INAF Osservatorio Astr. di Teramo, via Maggini, I-64100,
  Teramo, Italy \email{cantiello@oa-teramo.inaf.it}
  \and Astronomy Research Unit, University of Oulu, Pentti Kaiteran katu 1, 90014 Oulu, Finland
  \and INAF -- Osservatorio Astr. di Capodimonte Napoli, Salita Moiariello, 80131, Napoli, Italy
  \and Dip. di Fisica ``E. Pancini'', Universit\'a di Napoli Federico II, C.U. di Monte  Sant'Angelo, Via Cintia, 80126 Naples, Italy
  \and INFN, Sez. di Napoli, via Cintia, 80126, Napoli, Italy
  \and Center for Astrophysics | Harvard \& Smithsonian, 60 Garden Street, 02138 Cambridge (MA);
  \and European Southern Observatory, Karl-Schwarzschild-Str. 2, D-85748, Garching bei M\"unchen, Germany
  \and European Southern Observatory, Alonso de Cordova 3107, Vitacura, Santiago, Chile
  \and INAF -- Osservatorio Astr. di Arcetri, Largo Enrico Fermi 5, 50125, Firenze, Italy
  \and Kapteyn Astronomical Institute, University of Groningen, P.O. Box 72, 9700 AV Groningen, The Netherlands
  \and Instituto de Astrof\'sica de Canarias, Calle V\'ia L\'actea s/n, 38200 La Laguna, Tenerife, Spain
  \and Depto. Astrof\'isica, Universidad de La Laguna, Calle Astrof\'isico Francisco S\'anchez s/n, 38206 La Laguna, Tenerife, Spain
  \and School of Physics and Astronomy, Sun Yat-sen University, Zhuhai Campus, 2 Daxue Road, Xiangzhou District, Zhuhai, P. R. China
  \and Department of Astrophysics, University of Vienna T\"urkenschanzstra\ss e 17, 1180 Vienna, Austria}

   \date{Received ---; accepted ---}
 
  \abstract
  {A possible pathway for understanding the events and the mechanisms
    involved in galaxy formation and evolution is an in-depth
    comprehension of the galactic and inter-galactic fossil
    sub-structures with long dynamical
    times-scales: stars in the field and in stellar clusters.}
  {This paper continues the series of the Fornax Deep Survey (FDS).
    Following the previous studies dedicated to extended Fornax
    cluster members, in this paper we present the catalogs of compact
    stellar systems in the Fornax cluster as well as 
    extended background sources and point-like sources.}
 {We derive $ugri$ photometry of $\sim1.7$ million sources over
   the $\sim21$ square degree area of FDS centered on the bright central
   galaxy NGC\,1399. For a wider area, of $\sim27$ square degrees
   extending in the direction of NGC\,1316, we provide $gri$
   photometry for $\sim3.1$ million sources. To improve the
   morphological characterization of sources we generate multi-band
   image stacks by coadding the best seeing $gri$-band single
   exposures with a cut at FWHM$\leq0\farcs9$. We use the multi-band
   stacks as master detection frames, with a FWHM improved by
   $\sim15\%$ and a FWHM variability from field to field reduced by a
   factor of $\sim2.5$ compared to the pass-band with best FWHM, namely the $r$-band.
     The identification of compact sources, in particular of globular clusters (GC),
     is obtained from a combination of photometric (e.g. colors, magnitudes)
     and morphometric (e.g. concentration index, elongation, effective radius)
     selection criteria, by also taking as reference the properties
     of sources with well-defined classification from spectroscopic
     or high-resolution imaging data.}
   {Using the FDS catalogs, we present a preliminary analysis of
    globular cluster (GC) distributions in the Fornax area. The study
    confirms and extends further previous results which were limited
    to a smaller survey area. We observe the inter-galactic population
    of GCs, a population of mainly blue GCs centered on NGC\,1399,
    extends over $\sim0.9$ Mpc, with an ellipticity $\epsilon\sim0.65$
    and a small tilt in the direction of NGC\,1336. Several
    sub-structures extend over $\sim0.5~Mpc$ along various
    directions. Two of these structures do not cross any bright
    galaxy; one of them appears to be connected to NGC\,1404, a bright
    galaxy close to the cluster core and particularly poor of
    GCs. Using the $gri$ catalogs we analyze the GC distribution over
    the extended FDS area, and do not find any obvious GC
    sub-structure bridging the two brightest cluster galaxies, NGC\,1316
      and NGC\,1399. Although NGC\,1316 is more than twice brighter
      of NGC\,1399 in optical bands, using $gri$ data,
    we estimate a factor of $\sim3-4$ richer GC population
    around NGC\,1399 compared to NGC\,1316, out to galactocentric
    distances of $\sim40\arcmin$ or $\sim230$ kpc.}
  {The $ugri$ and $gri$ catalogs we present are made public via the
    FDS project web pages, and through the virtual observatory. Further
    studies, based on the catalogs, are in progress.}


\maketitle  
%

\section{Introduction}

The study of local complexes of galaxies --galaxy clusters and
groups-- is crucial for our understanding of the history of formation
and evolution of the Universe through its building blocks. Local
galaxy systems mark the endpoint of the evolution of galaxies after
billion years of more or less intense interaction with their companions
\citep[e.g.][]{mo10}.

A detailed study of the two extreme  structures in terms of
stellar density gives precious information on the history of formation
and interactions of a galaxy: faint extended stellar  features in
the outskirts of galaxies, characterized by low star density and very
long dynamical mixing time-scales \citep{johnston08}, and compact
stellar systems, which are intrinsically bright, have typically old
ages and have orbits that can trace recent and ancient accretion
  events \citep{brodie06}. The stratification of dense star clusters
and low surface brightness features can probe a galaxy environment on
different time scales, from the earliest epoch of formation to the
most recent merging events \citep[e.g.][]{west04,bournaud11}.

In the last decade, also thanks to the advent of efficient
large-format imaging cameras, many observational programs have carried
out intensive surveys dedicated to cover large fractions of nearby
galaxy systems, superseding in terms of both limiting magnitude and
spatial resolution any previous optical/near-IR study
\citep[e.g.][]{ferrarese12, iodice16}, and providing a rich variety of
data ideal for investigating compact stellar systems and faint stellar
structures in different galaxy environments
\citep[][]{dejong13,munoz14, durrell14, iodice19,venhola19,wittmann19}

In this framework, the Fornax Deep Survey, FDS, has surveyed the
\object{Fornax galaxy cluster} centered on \object{NGC\,1399} out to one virial radius,
and further extended observations in the direction of the Fornax
  A sub-cluster in the South-West with its brightest member,
  \object{NGC\,1316}, with a list of scientific topics: diffuse light and
intracluster medium \citep{iodice16}, galaxy scaling relations
\citep{iodice19, venhola17,venhola19, raj19}, extragalactic star
clusters and, more in general, compact stellar systems
\citep{dabrusco16, cantiello18vegas}, etc. In addition, the
  survey also contributes to research programs dealing with the study of
  the background galaxy population (e.g. identification of lensed
  systems and of their physical properties), and spectroscopic
  programs --for globular clusters \citep{pota18}, planetary nebulae
  \citep{spiniello18}, IFU study of galaxies in the cluster
  \citep{mentz16}.

The aim of this paper is to present the photometric and morphometric
catalog of all point-like and slightly extended sources of the survey,
describing the methodology used to characterize the sources. As key
topics of the survey, we present a preliminary study of compact
stellar systems, including globular clusters (GCs) and ultra compact
dwarf galaxies (UCDs).

Extragalactic, unresolved GCs are possibly the simplest class of
astrophysical objects beyond stars. To a first approximation, GCs host
a simple -- i.e., single age and single metallicity-- stellar
population. In spite of the results on multiple populations in
globular clusters \citep[e.g.][]{piotto07,carretta09,bastian18},
doubtless GCs host a stellar population much simpler than galaxies, in
terms of the metallicity and age distributions, because of the simpler
star-formation history which makes it possible to constrain the
properties of these systems at a higher level of precision with
respect to more complex and massive stellar systems.

The intrinsic simplicity of GCs, and of similar compact stellar
systems, together with the old ages and the high luminosity, make
these astronomical sources powerful and robust tracers of a galaxy and
its environment, suitable to study a galaxy and its relevant
structures out to cosmological distances \citep{alamo13, janssens17,
  vanzella17}. The rich set of observables of stellar clusters
  makes them useful fossil records of the history of the evolution of
  their host galaxy and indicators of some of its physical property
  (distance, merging history, mass, metallicity, etc.).  Here we
focus on preliminary projected distribution maps of GCs and UCDs, and
postpone further analysis of these sources to a forthcoming paper
(Cantiello et al., 2020, in prep.).

In what follows we will assume a distance modulus of $(m{-}M) = 31.51
\pm 0.03~(ran.) \pm 0.15~(sys.)$ mag for the Fornax galaxy cluster,
corresponding to $d = 20.0 \pm 0.3~(ran.) \pm 1.4~(sys.)$ Mpc
\citep{blake09}.

The paper is organized as follows. In section \ref{data} we describe
the data, the procedures for source identification, calibration and
characterization, and present the final FDS catalog of compact and
slightly extended sources, and background galaxies. Section \ref{maps}
is dedicated to a pilot usage of the catalogs, for deriving 2-D
distributions of compact sources in the area. In Section
\ref{ctortora} the application to a science case for background
sources is reported. A brief summary of our conclusions is presented
in Section \ref{summary}.


\section{Data and data analysis}
\label{data}
\subsection{Observations and data reduction}

The observations used in this work are part of the now completed FDS survey.

The FDS consists of a combination of Guaranteed Time Observations from
the Fornax Cluster Ultra-deep Survey (FOCUS, P.I. R. Peletier) and the
VST Early-type GAlaxy Survey (VEGAS, P.I.  E. Iodice). The surveys are
both performed with the ESO VLT Survey Telescope (VST), which is a
2.6-m diameter optical survey telescope located at Cerro Paranal,
Chile \citep{schipani10}.  The imaging is in the $u$, $g$,$r$ and
$i$-bands using the $1\times 1$ square degree field of view camera
OmegaCAM \citep{kuijken11}.

\begin{figure*}
   \centering
    \includegraphics[trim={1cm 0cm 0.2cm 0cm},clip,width=9.15cm]{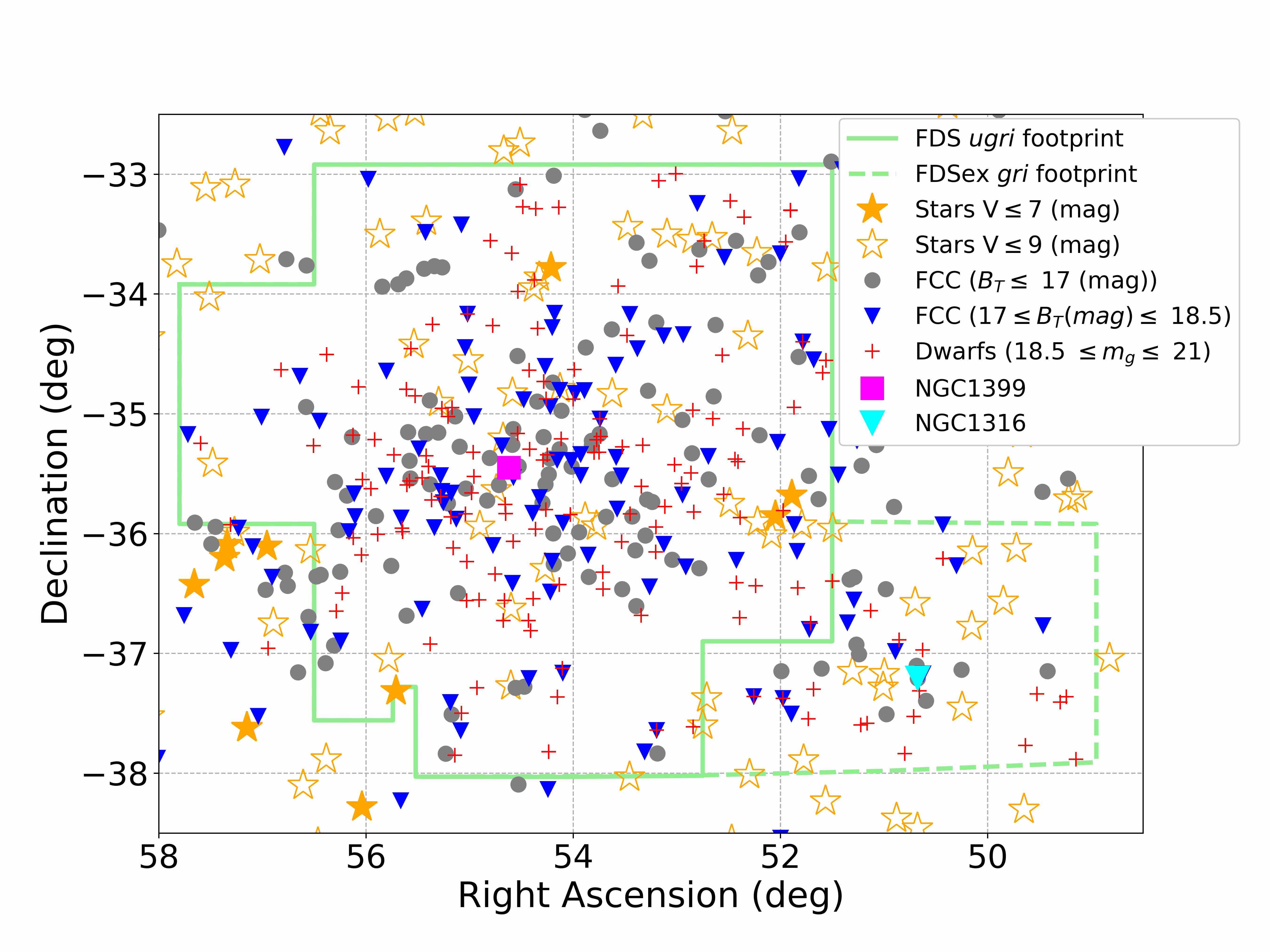}
    \includegraphics[trim={1cm 0cm 0.2cm 0cm},clip,width=9.15cm]{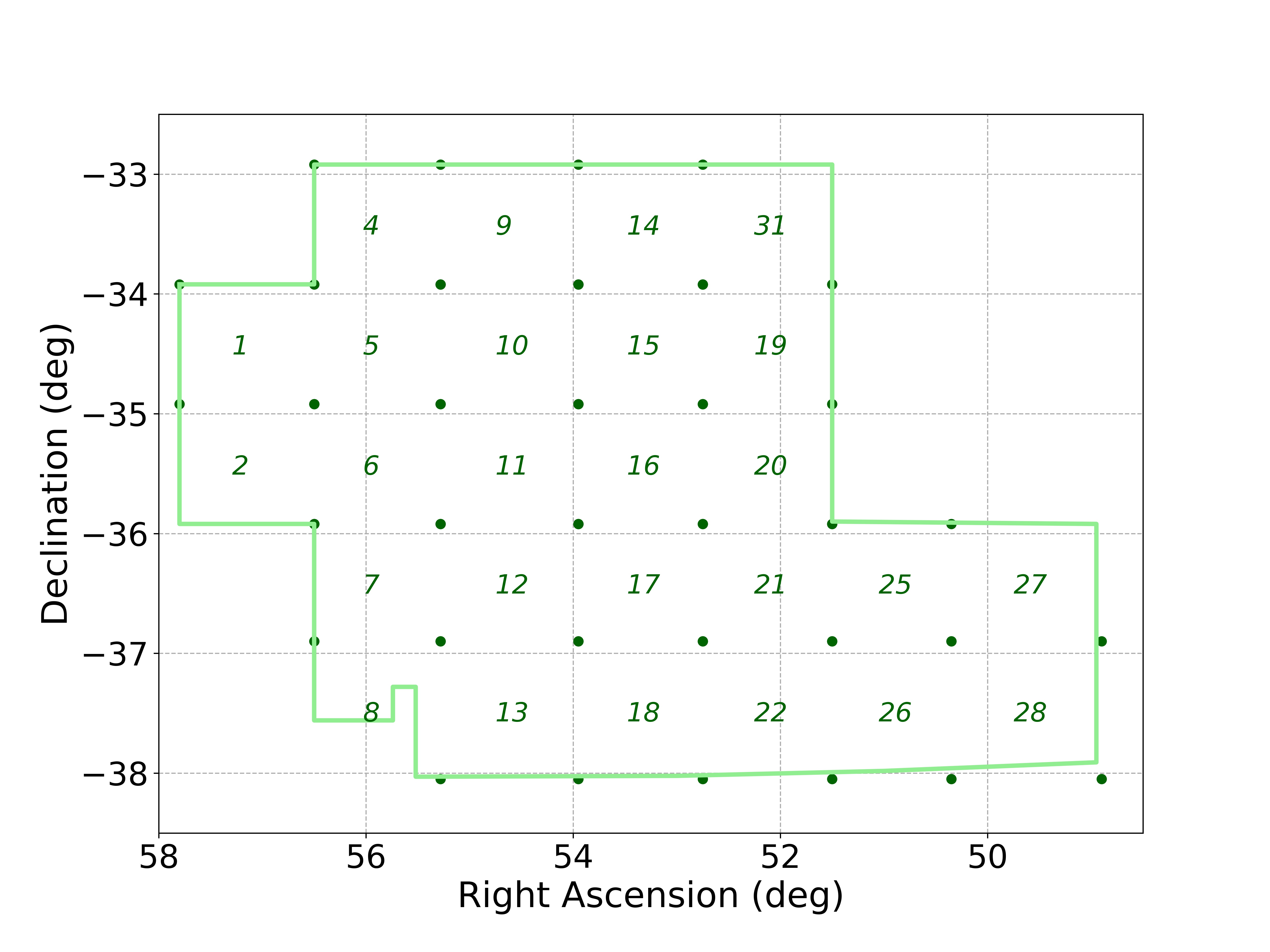}
  \caption{Left panel: The FDS footprint of the area covered by $ugri$
    photometry (green solid line), and by only $gri$ (dashed green
    line). Other sources from catalogs available in the literature are
    also shown, as labeled. Bright galaxies from the Fornax Cluster
    Catalog \citep{ferguson89} are subdivided into two
      categories: likely members brighter than $B_T=17$ mag and with
      $17\leq B_T~(mag)\leq18.5$ \citep[filled gray circles and blue
        triangles, respectively; from][Table II]{ferguson89}. Dwarf
      galaxies from FDS by \citet{venhola18}, in the magnitude range
      $18.5 \leq m_g~(mag)\leq 21$, are indicated with red
      crosses. The positions of the two brightest galaxies, NGC\,1316
    and NGC\,1399, are also shown with a filled cyan triangle and a
    magenta square, respectively. Orange filled/empty five pointed
    stars mark stars with $m_V\leq7/\leq9$ mag, respectively. Right
    panel: The FDS and FDSex area. Green lines mark the edges of the
    survey, green bullets show the edges of single pointings; the ID
    of the field is also indicated.  }
              \label{anicemap}%
    \end{figure*}

The main body of the FDS dataset is centered on NGC\,1399, the second
brightest galaxy of the Fornax galaxy cluster in optical bands
  and the brightest galaxy of the main cluster, and consists of 21
VST fields with a complete $ugri$ coverage.  Further five fields in
the $gri$ bands extend in the south-west direction of the cluster, the
Fornax A sub-cluster which cover the regions of the brightest cluster
galaxy, the peculiar elliptical NGC\,1316. For sake of clarity, in
what follows we refer to the 21 FDS fields with $ugri$ as FDS survey,
and to the entire sample of 26 fields with $gri$ coverage as
FDS-extended, or FDSex. The FDS and FDSex areas are shown in Figure
\ref{anicemap}; some of the known objects available from the
literature and from previous FDS works are marked in the left panel of
the figure.

The data, data acquisition and reduction procedures are presented in a
number of papers of the FDS series \citep{iodice16, iodice17a,
  iodice17b, iodice19, venhola17, venhola18, venhola19}. A full
  description of the observations and the pipeline used for data
  reduction \citep[AstroWISE,][]{mcfarland13} steps are given in the
  cited papers, and in Peletier et al. (in prep.).

In what follows we describe two critical differences with
respect to previous works, related to the focus on compact stellar
  systems in the present work.

\subsection{Multi-band image stacks}
\label{a-stacks}

\begin{figure*}
   \centering
   \includegraphics[trim={2.5cm 0cm 2.5cm 0.5cm},clip,width=\textwidth]{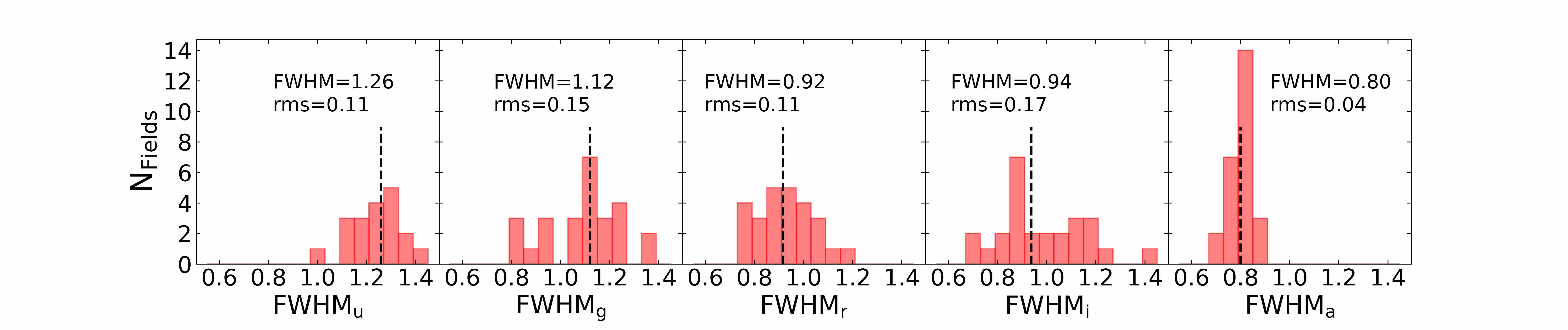}
   \caption{Histograms of the median PSF FWHM of the FDS fields in the
     four available passbands, plus the multi-band $a$-stacks. The
     vertical dashed line shows the median of the ensemble. }
              \label{histofwhm}%
    \end{figure*}

The FDS standard reduction pipeline produced imaging data for many
different scientific cases, with a general focus on extended galaxies
in the cluster \citep[e.g.][]{spavone17}. In columns (2-5) of Table
\ref{tab_iq}, we report the median full width at half maximum, FWHM,
of the point spread function, PSF, in arcseconds for each FDS VST
field and for each available band; the FWHM distributions are also
shown in the histograms of Figure \ref{histofwhm}.  The large FWHM
variation, up to $\sim50\%$ for different fields observed in the same
passband, may represent a limitation to the effectiveness of the FDS
dataset for the science cases related to compact objects (foreground
MW stars, background galaxies, GCs host in Fornax, etc.). The typical
FWHM scatter of the exposures combined to obtain the single FDS fields
stacks is $rms_{MAD}=0\farcs36,~0\farcs21,~0\farcs33,~0\farcs21$ in
$u/g/r/i$-band, respectively.

To improve the detection and characterization of compact sources, we
combined in a single coadded image all single VST exposures in $g$,
$r$ and $i$ bands with a median FWHM lower than a fixed upper limit,
$u$-band exposures were ignored because of the lower signal-to-noise
and worse FWHM. After various experiments, we fixed the FWHM limit to
$0\farcs9$: if a lower FWHM cut is adopted, the final resolution of
the stack improves, at the expenses of a worse detection limit and
larger field-to-field mean FWHM variability; a higher FWHM cut,
instead, would make ineffective the use of multi-band stacks compared
to single bands images. Hence, the $0\farcs9$ cut is adopted as the
trade off between needs of better resolution and uniformity of the
master detection frame. The combined image was processed as the single
band images, except for the photometric calibration which is not
derived.  In the following, we refer to the coadd of $gri$ exposures
with FWHM cut as $a$-stack, and use the subscript $a$ to identify the
quantities derived from it.  With this procedure, a new frame with
narrower and more stable FWHM compared with $ugri$ bands is obtained,
and used as master detection frame. This improved both the uniformity
of detections over the different FDS fields, and the determination of
the morphological properties of the sources, allowing more accurate
characterization of compact and point-like objects. These $a$-stacks
will not be used to define absolute quantities (like calibrated
magnitudes), but only for relative ones (like the $CI_n$, see below),
thus the wavelength dependence of the PSF and source morphology will
not be an issue.

As shown in Table \ref{tab_iq}, the $a$-stacks have a median FWHM
smaller by $\sim15\%$ and with an $rms$ scatter a factor of $\sim2.5$
lower than the median and $rms$ of the FWHM for the best passband,
namely the $r$-band.

\begin{figure*}
   \centering
    \includegraphics[width=\textwidth]{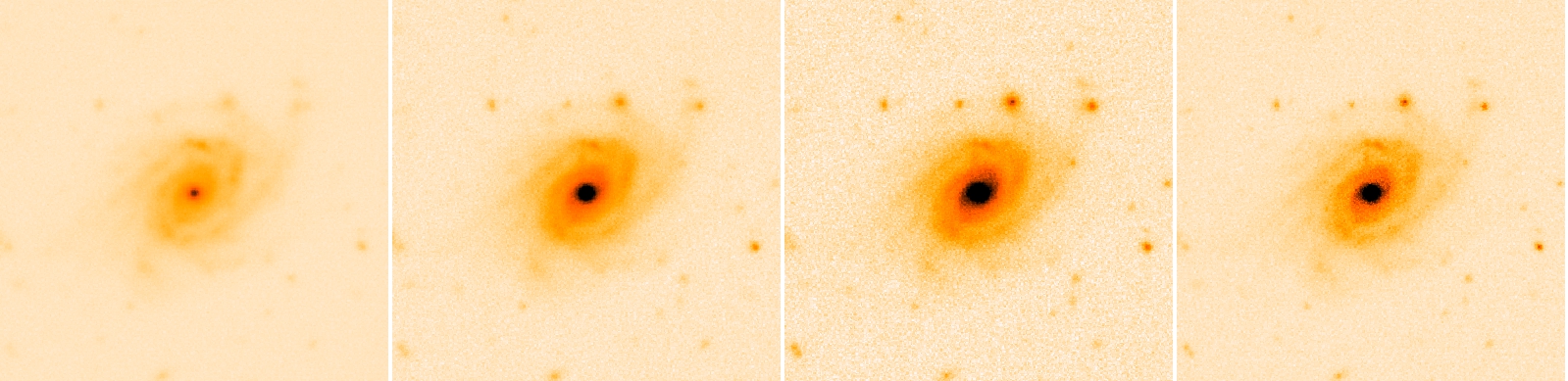}
   \caption{From left to right: $g$, $r$, $i$-band and $a$-stack of a
     background spiral galaxy in the field FDS\#5 \citep[$FCCB\,1532$,
     ][]{ferguson89}. The rightmost panel is derived from the
     combination of the sub-exposures of the first three panels,
     selecting only the ones with lowest atmospheric turbulence (see
     text).}
              \label{fds05}%
    \end{figure*}

In Figure \ref{fds05} we show a $1\arcmin \times1\arcmin$ thumbnail of
the same FDS region in $g$, $r$, and $i$-band and the $a$-stack image
centered on background spiral galaxy in the field FDS\#5
\citep[$FCCB\,1532$, ][]{ferguson89}. In general, the depth of the
coadded multiband $a$-stack does not change much compared with the
best band of the field, because the reduced number of exposures used
is compensated by the better S/N due to the higher spatial resolution.
The spatial resolution, however, is in all cases enhanced, as shown in
the $FWHM_a$ column in Table \ref{tab_iq}.

\begin{sidewaystable*}
\tiny
  \caption{Image quality parameters for FDS and FDSex fields}
\label{tab_iq}      
\centering                          
\begin{tabular}{c c c c c c |c c |c c |c c| c c c c }        
\hline\hline                 
Field ID & $FWHM_u$ & $FWHM_g$   & $FWHM_r$     & $FWMH_i$    & $FWHM_a$   &$\langle P_2(s)\rangle $&$\sigma[P_2(s)]$ &$\langle P_2(w)\rangle $&$\sigma[P_2(w)]$&$\langle P_2(x)\rangle $&$\sigma[P_2(x)]$ & $u_{lim}$ & $g_{lim}$ &$r_{lim}$ &$i_{lim}$ \\ 
         & $(\arcsec)$        &  $(\arcsec)$         &    $(\arcsec)$         &  $(\arcsec)$         &  $(\arcsec)$          &  (mag)   &  (mag )&  (mag)     &  (mag)  &  (mag)   & (mag)&  (mag)     &  (mag)  &  (mag)   & (mag)                \\          
  (1)    &      (2)           &       (3)            &        (4)             &       (5)            &       (6)             &  (7)     &  (8)   &  (9)       &  (10)   &  (11)    & (12) &  (13)      &  (14)   &  (14)    & (16)                 \\          
\hline                       
   1 &     1.17 $\pm$   0.03 &     1.35 $\pm$   0.12 &     1.14 $\pm$   0.11 &     0.69 $\pm$   0.08 &     0.72 $\pm$  0.08  &   0.006  & 0.019  &     0.018  & 0.024   &   0.014  & 0.030  &  24.24$\pm$ 0.13  &  25.39$\pm$ 0.10   &    24.65 $\pm$0.17    &   24.53 $\pm$0.15 \\
   2 &     1.21 $\pm$   0.08 &     1.11 $\pm$   0.16 &     0.89 $\pm$   0.05 &     0.79 $\pm$   0.09 &     0.79 $\pm$  0.04  &   0.008  & 0.018  &     0.011  & 0.020   &   0.005  & 0.032  &  24.02$\pm$ 0.18  &  25.41$\pm$ 0.13   &    25.04 $\pm$0.12    &   24.12 $\pm$0.13 \\
   4 &     1.19 $\pm$   0.05 &     1.39 $\pm$   0.13 &     1.19 $\pm$   0.09 &     0.70 $\pm$   0.07 &     0.71 $\pm$  0.07  &   0.007  & 0.020  &     0.052  & 0.026   &  -0.005  & 0.029  &  24.12$\pm$ 0.09  &  25.35$\pm$ 0.10   &    24.65 $\pm$0.11    &   24.44 $\pm$0.14 \\
   5 &     1.35 $\pm$   0.07 &     1.17 $\pm$   0.14 &     0.98 $\pm$   0.06 &     1.11 $\pm$   0.16 &     0.82 $\pm$  0.09  &   0.005  & 0.018  &     0.007  & 0.024   &   0.012  & 0.033  &  24.05$\pm$ 0.17  &  25.48$\pm$ 0.10   &    24.72 $\pm$0.08    &   23.88 $\pm$0.10 \\
   6 &     1.13 $\pm$   0.07 &     0.83 $\pm$   0.04 &     1.08 $\pm$   0.12 &     1.24 $\pm$   0.14 &     0.80 $\pm$  0.05  &   0.005  & 0.021  &     0.023  & 0.023   &  -0.007  & 0.033  &  24.22$\pm$ 0.10  &  25.70$\pm$ 0.10   &    24.66 $\pm$0.14    &   23.51 $\pm$0.09 \\
   7 &     1.03 $\pm$   0.06 &     0.82 $\pm$   0.04 &     0.90 $\pm$   0.07 &     1.44 $\pm$   0.13 &     0.78 $\pm$  0.08  &   0.007  & 0.022  &     0.012  & 0.020   &   0.008  & 0.026  &  24.16$\pm$ 0.12  &  25.79$\pm$ 0.11   &    24.91 $\pm$0.13    &   23.35 $\pm$0.11 \\
   8 &     1.21 $\pm$   0.10 &     0.93 $\pm$   0.16 &     0.90 $\pm$   0.12 &     0.96 $\pm$   0.18 &     0.84 $\pm$  0.13  &   0.006  & 0.021  &     0.022  & 0.024   &  -0.015  & 0.033  &  24.23$\pm$ 0.21  &  25.50$\pm$ 0.22   &    24.99 $\pm$0.21    &   23.72 $\pm$0.20 \\
   9 &     1.42 $\pm$   0.08 &     1.20 $\pm$   0.07 &     0.97 $\pm$   0.08 &     0.84 $\pm$   0.07 &     0.82 $\pm$  0.05  &   0.003  & 0.022  &     0.036  & 0.021   &  -0.015  & 0.035  &  23.96$\pm$ 0.09  &  25.50$\pm$ 0.12   &    25.06 $\pm$0.13    &   24.38 $\pm$0.13 \\
  10 &     1.34 $\pm$   0.06 &     1.15 $\pm$   0.05 &     0.96 $\pm$   0.07 &     1.09 $\pm$   0.11 &     0.86 $\pm$  0.05  &   0.000  & 0.018  &     0.009  & 0.014   &  -0.010  & 0.026  &  24.09$\pm$ 0.10  &  25.52$\pm$ 0.09   &    24.84 $\pm$0.11    &   23.96 $\pm$0.11 \\
  11 &     1.27 $\pm$   0.06 &     1.09 $\pm$   0.14 &     1.08 $\pm$   0.14 &     1.17 $\pm$   0.10 &     0.84 $\pm$  0.08  &   0.011  & 0.023  &     0.025  & 0.024   &  -0.002  & 0.032  &  24.09$\pm$ 0.13  &  25.22$\pm$ 0.10   &    24.65 $\pm$0.11    &   23.64 $\pm$0.09 \\
  12 &     1.18 $\pm$   0.08 &     0.80 $\pm$   0.04 &     0.97 $\pm$   0.07 &     1.17 $\pm$   0.09 &     0.80 $\pm$  0.05  &   0.014  & 0.024  &     0.022  & 0.021   &   0.001  & 0.037  &  24.30$\pm$ 0.09  &  25.74$\pm$ 0.11   &    24.85 $\pm$0.11    &   23.61 $\pm$0.09 \\
  13 &     1.10 $\pm$   0.05 &     0.91 $\pm$   0.05 &     1.03 $\pm$   0.08 &     1.16 $\pm$   0.07 &     0.89 $\pm$  0.05  &   0.003  & 0.016  &     0.021  & 0.016   &  -0.003  & 0.029  &  24.39$\pm$ 0.18  &  25.72$\pm$ 0.13   &    24.99 $\pm$0.12    &   24.22 $\pm$0.13 \\
  14 &     1.46 $\pm$   0.09 &     1.18 $\pm$   0.09 &     0.96 $\pm$   0.08 &     0.85 $\pm$   0.06 &     0.83 $\pm$  0.06  &   0.004  & 0.019  &     0.012  & 0.017   &   0.005  & 0.028  &  23.99$\pm$ 0.08  &  25.43$\pm$ 0.12   &    24.94 $\pm$0.13    &   24.30 $\pm$0.12 \\
  15 &     1.30 $\pm$   0.05 &     1.13 $\pm$   0.04 &     0.88 $\pm$   0.04 &     0.98 $\pm$   0.09 &     0.81 $\pm$  0.06  &   0.001  & 0.020  &     0.008  & 0.022   &  -0.001  & 0.031  &  24.19$\pm$ 0.11  &  25.37$\pm$ 0.09   &    25.10 $\pm$0.08    &   24.02 $\pm$0.16 \\
  16 &     1.31 $\pm$   0.04 &     1.26 $\pm$   0.08 &     0.91 $\pm$   0.08 &     1.09 $\pm$   0.07 &     0.84 $\pm$  0.05  &   0.008  & 0.025  &     0.006  & 0.020   &  -0.000  & 0.035  &  24.16$\pm$ 0.11  &  25.31$\pm$ 0.08   &    24.93 $\pm$0.11    &   23.88 $\pm$0.09 \\
  17 &     1.27 $\pm$   0.06 &     1.25 $\pm$   0.16 &     0.82 $\pm$   0.05 &     1.01 $\pm$   0.07 &     0.80 $\pm$  0.04  &  -0.006  & 0.020  &     0.020  & 0.020   &  -0.011  & 0.032  &  24.17$\pm$ 0.09  &  25.16$\pm$ 0.18   &    25.21 $\pm$0.11    &   24.01 $\pm$0.10 \\
  18 &     1.12 $\pm$   0.08 &     0.94 $\pm$   0.05 &     1.03 $\pm$   0.07 &     1.12 $\pm$   0.12 &     0.87 $\pm$  0.09  &  -0.002  & 0.018  &     0.021  & 0.016   &   0.007  & 0.025  &  24.19$\pm$ 0.23  &  25.57$\pm$ 0.13   &    24.93 $\pm$0.12    &   24.14 $\pm$0.13 \\
  19 &     1.26 $\pm$   0.05 &     1.14 $\pm$   0.09 &     0.89 $\pm$   0.05 &     0.86 $\pm$   0.06 &     0.79 $\pm$  0.05  &   0.010  & 0.022  &     0.042  & 0.022   &   0.009  & 0.025  &  24.10$\pm$ 0.11  &  25.46$\pm$ 0.09   &    25.15 $\pm$0.10    &   24.13 $\pm$0.13 \\
  20 &     1.30 $\pm$   0.05 &     1.23 $\pm$   0.06 &     0.92 $\pm$   0.09 &     1.08 $\pm$   0.08 &     0.81 $\pm$  0.08  &   0.019  & 0.033  &    -0.006  & 0.032   &   0.002  & 0.044  &  24.12$\pm$ 0.11  &  25.17$\pm$ 0.12   &    24.76 $\pm$0.13    &   23.80 $\pm$0.12 \\
  21 &     1.22 $\pm$   0.05 &     1.12 $\pm$   0.06 &     0.78 $\pm$   0.05 &     0.88 $\pm$   0.08 &     0.78 $\pm$  0.04  &   0.001  & 0.022  &     0.004  & 0.027   &   0.002  & 0.035  &  24.06$\pm$ 0.11  &  25.22$\pm$ 0.09   &    24.92 $\pm$0.12    &   24.22 $\pm$0.09 \\
  22 &               \nodata &     1.03 $\pm$   0.06 &     0.80 $\pm$   0.06 &     0.85 $\pm$   0.05 &     0.79 $\pm$  0.05  &  \nodata & \nodata&     0.004  & 0.019   &   0.007  & 0.029  &    \nodata        &  25.27$\pm$ 0.14   &    24.92 $\pm$0.13    &   24.21 $\pm$0.10 \\
  25 &               \nodata &     1.12 $\pm$   0.05 &     0.76 $\pm$   0.06 &     0.85 $\pm$   0.08 &     0.78 $\pm$  0.08  &  \nodata & \nodata&     0.016  & 0.025   &  -0.003  & 0.031  &    \nodata        &  25.36$\pm$ 0.11   &    24.98 $\pm$0.12    &   24.10 $\pm$0.10 \\
  26 &               \nodata &     0.95 $\pm$   0.12 &     0.80 $\pm$   0.04 &     0.91 $\pm$   0.08 &     0.78 $\pm$  0.04  &  \nodata & \nodata&     0.037  & 0.021   &  -0.018  & 0.035  &    \nodata        &  24.79$\pm$ 0.15   &    25.00 $\pm$0.11    &   23.94 $\pm$0.13 \\
  27 &               \nodata &     1.05 $\pm$   0.09 &     0.78 $\pm$   0.06 &     0.89 $\pm$   0.09 &     0.77 $\pm$  0.08  &  \nodata & \nodata&     0.012  & 0.026   &  -0.007  & 0.042  &    \nodata        &  25.19$\pm$ 0.11   &    24.79 $\pm$0.15    &   23.75 $\pm$0.16 \\
  28 &               \nodata &     1.09 $\pm$   0.07 &     0.79 $\pm$   0.15 &     0.91 $\pm$   0.11 &     0.78 $\pm$  0.10  &  \nodata & \nodata&     0.007  & 0.035   &   0.013  & 0.032  &    \nodata        &  25.11$\pm$ 0.13   &    24.65 $\pm$0.19    &   23.78 $\pm$0.14 \\
  31 &    1.46 $\pm$    0.08 &     1.22 $\pm$   0.06 &     1.00 $\pm$   0.07 &     0.86 $\pm$   0.08 &     0.84 $\pm$  0.05  &   0.012  & 0.022  &     0.030  & 0.023   &   0.003  & 0.032  &  23.83$\pm$ 0.11  &  25.31$\pm$ 0.11   &    24.78 $\pm$0.15    &   24.05 $\pm$0.17 \\   
\hline                                                                                                                                                                                                                                                                         
{\tiny Median} & 1.26$\pm$ 0.11 & 1.12$\pm$   0.15   &     0.92 $\pm$   0.11 & 	   0.94 $\pm$   0.17 &	   0.80  $\pm$ 0.04  &   0.006  & 0.021  &     0.017  & 0.022   &   0.000  & 0.032 &  24.12$\pm$ 0.13  &  25.38$\pm$ 0.17   &    24.92 $\pm$0.17    &   24.02 $\pm$0.24 \\  
\hline                                   
\end{tabular}
\end{sidewaystable*}

\subsection{Photometry and photometric calibration}

Catalogs are derived for each single FDS pointing; the identification
of fields with available data is shown in the right panel of Figure
\ref{anicemap}.  To increase the contrast of faint sources close to
the cores of extended galaxies, before running the procedures to
obtain the photometry and the morphometry (like FWHM, elongation, flux
radius, etc.; see section \ref{morphometry} below) we modeled and
subtracted all Fornax members brighter than $B_T\sim18$ mag.  The fit
of the isophotes is performed using the IRAF STSDAS task ELLIPSE,
which is based on an algorithm by \citet{jedrzejewski87}.

To obtain the photometry of sources in FDS frames, we used a
combination of procedures, based on SExtractor \citep{bertin96} and
DAOphot \citep{stetson87} runs, and codes developed by the first
  author. We adopt AB mag photometric system, as in previous FDS
  works. The galaxy-subtracted frames used in this stage are already
calibrated as described in the previous works of the FDS series (see
below).

First, we used SExtractor to obtain the mean properties of each frame,
like the FWHM; the reference morphometry for each source is obtained
from the $a$-stacks, though we also derived the morphometric
properties for all available passbands. Then, DAOphot is run on the
$a$-stacks, and fed to our procedure to identify bright, non-saturated
and isolated stars needed to obtain a variable PSF model over the
single pointing. Typically, with this procedure we selected $\sim200$
candidate PSFs per single FDS field, that were visually inspected in
all bands to remove candidates contaminated by faint companions,
bright halos of galaxies or saturated stars, or other instrumental
artifacts. Using this iterative process, we ended up with a typical
list of 50 to 100 point-like sources to model the PSF with DAOphot for
each filter and field. The list of PSFs was then fed to DAOphot for
PSF modeling, adopting the variable PSF option. The first complete
DAOphot run was on the $a$-stack. The output table for this run was
used to $i)$ identify sources to define a master detection catalog,
$ii)$ obtain the DAOphot sharpness parameter, that will be used as
additional parameter for selecting good candidate compact sources.

The master detection catalog was then given as input to run DAOphot on
each available filter and for all fields: $ugri$ for the FDS
area, $gri$ for FDSex.

We also run SExtractor on the full set of images, to obtain the
aperture magnitude within 8-pixel diameter ($MAG\_APER$) and the
automated aperture magnitude derived from \citet{kron80} first moment
algorithms ($MAG\_AUTO$), with the respective photometric
errors\footnote{For SExtractor runs, we adopted Gaussian convolution
  kernels of different sizes depending on the FWHM of the field.}.
For the aperture magnitudes, after some tests we adopted the 8
  pixel diameter: larger diameters implied larger statistical errors
  on derived magnitudes (because of the noisier background and higher
  contamination from neighboring sources), smaller diameters suffered
  from larger systematic errors (because larger aperture corrections
  are needed). Both $MAG\_APER(8)$ and $MAG\_AUTO$ are stored in our
final catalogs; in particular $MAG\_AUTO$ provides a good choice for
the magnitude of non-compact background objects.

The photometric calibration is carried out in two steps. The first is
the same described in \citet{venhola18}, and uses standard star fields
observed each night and comparing their OmegaCAM magnitudes with the
final data from the Sloan Digital Sky Survey Data III \citep{alam15}.

With such calibration, and after applying the field and pass-band
dependent aperture corrections, the photometry of the same sources in
different adjacent FDS pointings show a spatially variable offset,
with a median upper limit of $\sim0.1$ mag. This might be a
consequence of the different (mean) photometric conditions for
neighboring FDS fields during the FDS observing runs which span a time
interval of $\sim5$ years.

As a second step of the photometric calibration, to improve the
photometric uniformity and consistency over the FDS (and FDSex) area,
and to derive the spatially and filter dependent aperture correction
map, we compared our VST photometry of bright non-saturated point-like
sources to the APASS photometry\footnote{Visit the URL
  https://www.aavso.org/} and obtained the two-dimensional map that
best matches the two datasets. The map is derived for each field
separately, using a support vector machine (SVM) supervised learning
method, with a radial basis function (RBF) kernel
\citep{pedregosa12}. Only isolated unsaturated stars, brighter than a
given magnitude cut (19/17/17/16.5 mag in $u/g/r/i$ band,
respectively) are used in the regression algorithm.

\begin{figure*}
   \includegraphics[trim={0cm 0cm 0cm 0.5cm},clip,width=\textwidth]{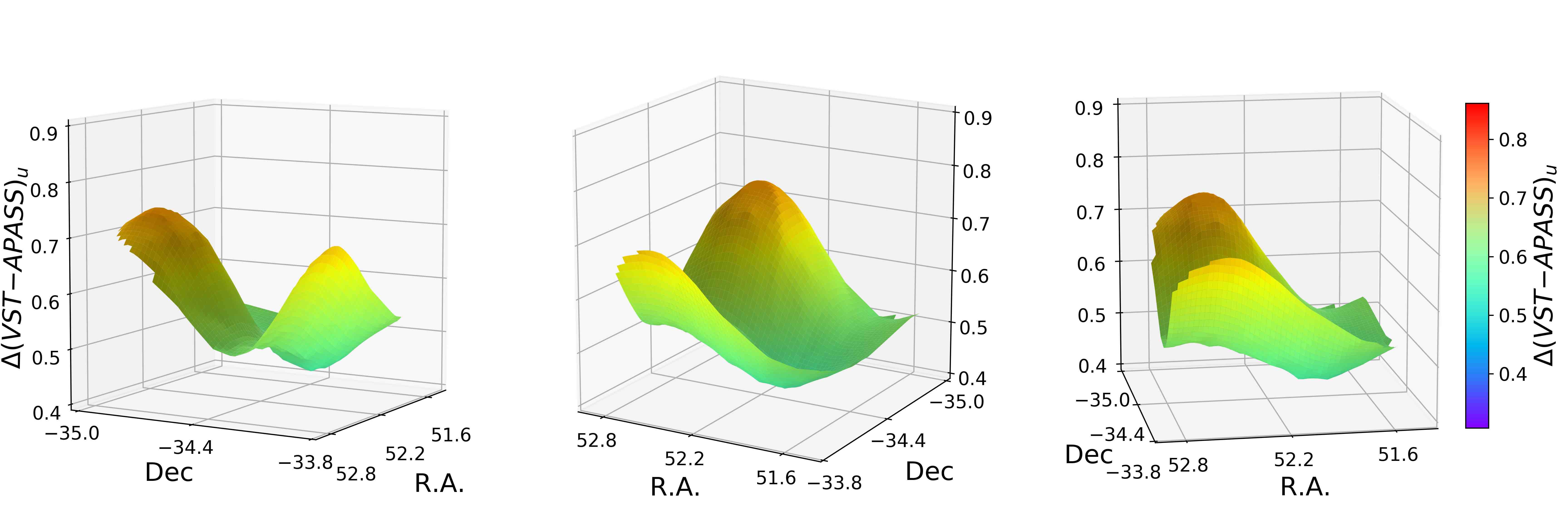}
   \vskip -0.25cm
   \includegraphics[trim={0cm 0cm 0cm 0.5cm},clip,width=\textwidth]{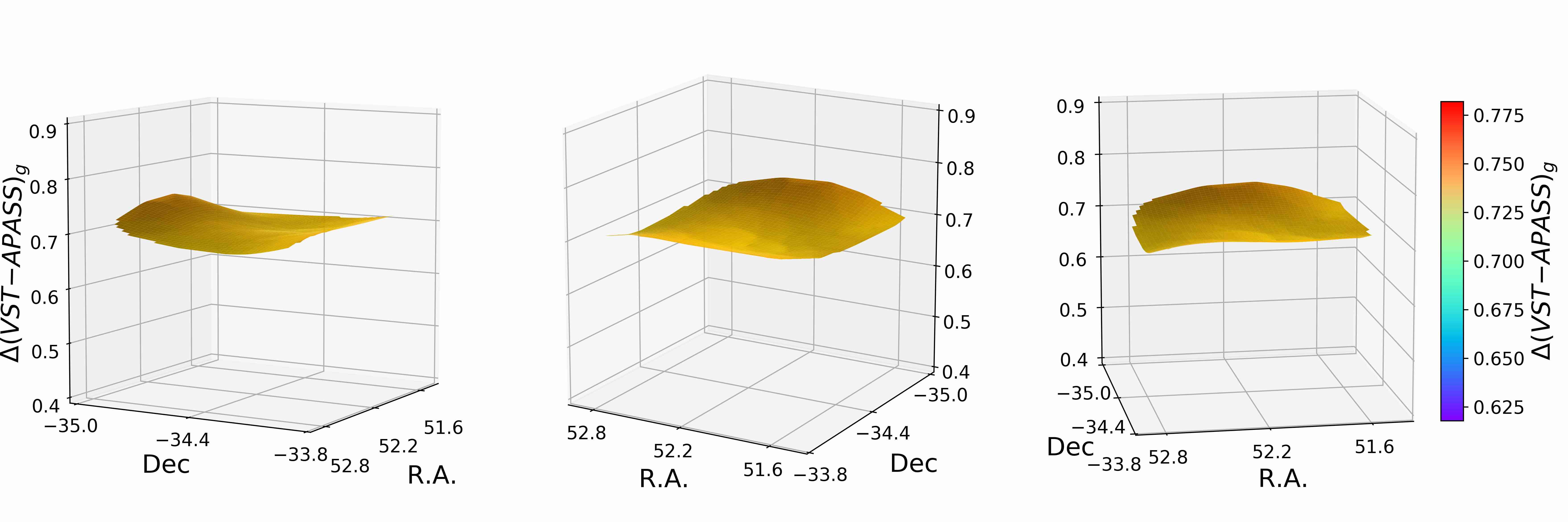}
   \vskip -0.25cm
   \includegraphics[trim={0cm 0cm 0cm 0.5cm},clip,width=\textwidth]{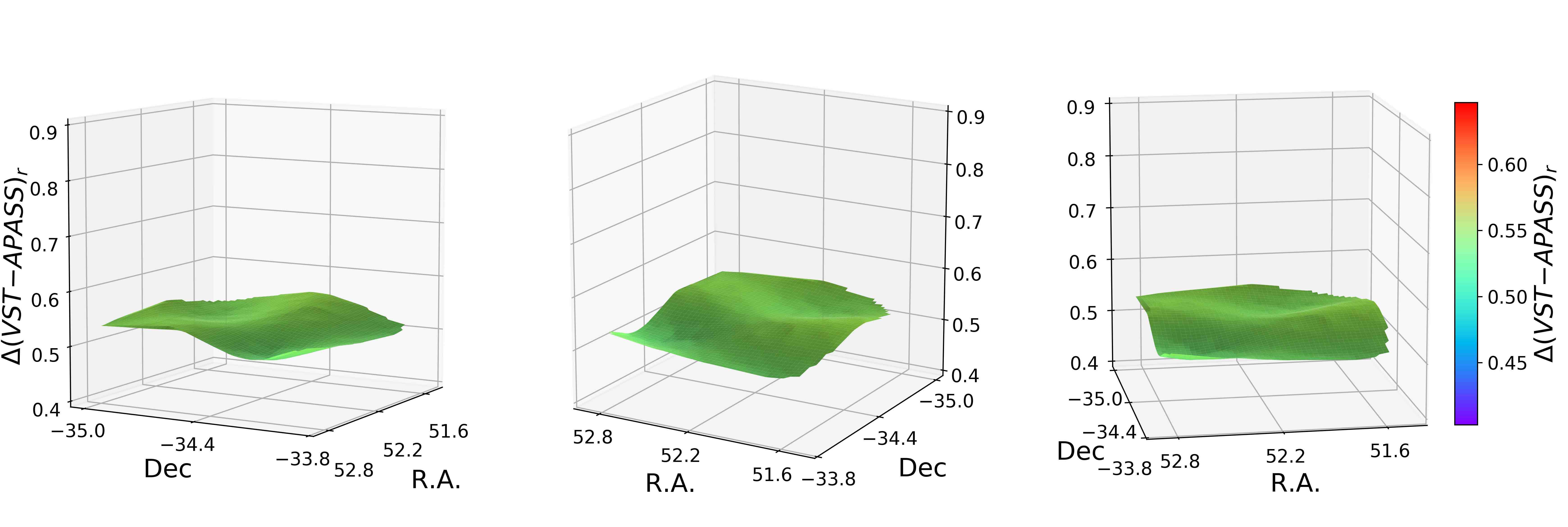}
   \vskip -0.25cm
   \includegraphics[trim={0cm 0cm 0cm 0.5cm},clip,width=\textwidth]{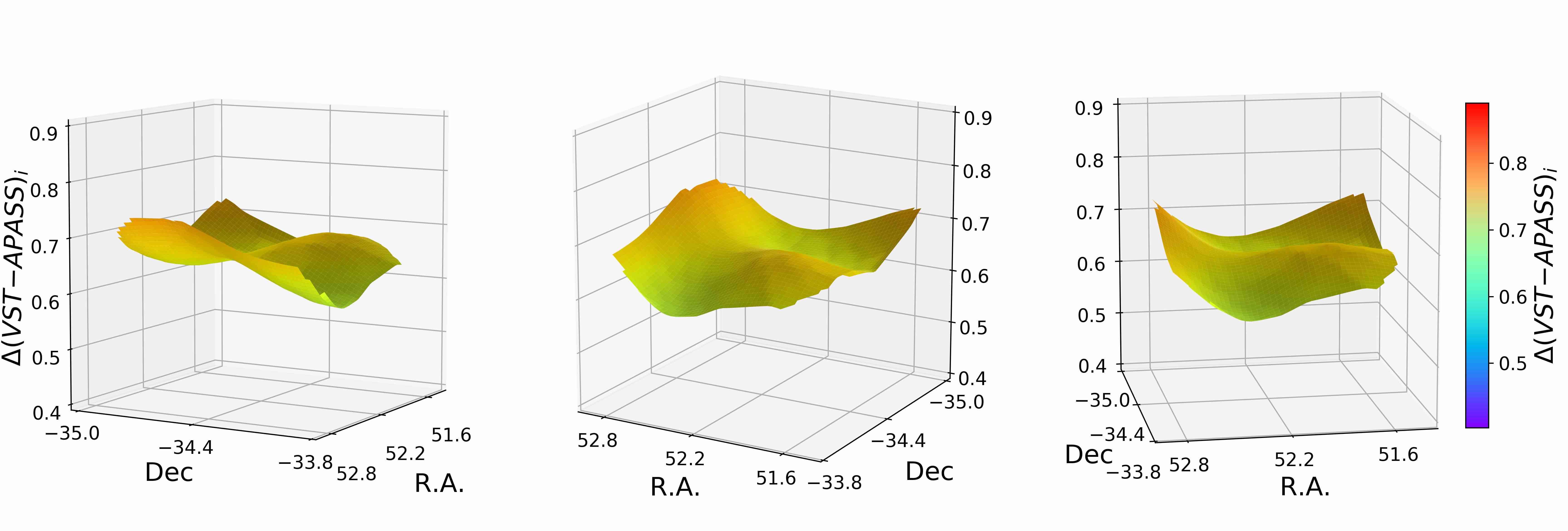}
      \caption{An example of the two-dimensional photometric
        correction maps for refining the photometry of FDS fields.
        The maps also include the aperture correction term.  Field
        FDS\#19 is shown: $u$, $g$, $r$ and $i$-band correction maps
        are plotted from upper to lower panels, respectively. For each
        passband the surface correction map is shown with the same
        color coding and for different viewpoints in each of the three
        panels.  }
         \label{f19_calibmap}
\end{figure*}

The correction maps are derived from 200 to 300 stars per FDS
  field, the final median $rms_{VST-APASS}$ between VST and APASS
photometry over the full set of re-calibrated frames is reported in
Table \ref{tab_vstsm}. Figure \ref{f19_calibmap} shows an example of
the correction maps derived for the field FDS\#19.  Each correction
map is then applied to its specific field and passband, to correct the
photometry of all sources detected in the specific FDS pointing.

Because APASS lacks $u$ coverage, for such passband we adopted a
slightly different re-calibration strategy. After the preliminary
calibration described above, the $B$-band magnitudes of stars from
APASS were transformed to $u$-band using Lupton (2005) transformation
equations available from the SDSS web
pages\footnote{https://www.sdss.org/dr12/algorithms/sdssUBVRITransform.php,
  Lupton (2005)}.  In particular:
$u=B_{APASS}+0.8116\cdot(u{-}g)_{fit}-0.1313$, where the
$(u{-}g)_{fit}$ color index is derived from the APASS $(g{-}i)$ and
$(g{-}r)$ indices, using a second degree polynomial fit derived from
SDSS data over different sky regions\footnote{The fitted relation is:
  $(u{-}g)_{fit}=P_{00}+\gi_{APASS}\times P_{10}+\gr_{APASS}\times
  P_{01}+\gi_{APASS}^2\times P_{20}+\gi_{APASS}\times
  \gr_{APASS}\times P_{11}+\gr_{APASS}^2\times P_{02}$, with $P_{00} =
  0.1997$, $P_{10} = -0.1799$, $P_{01} = 2.849$, $P_{20} = 1.043$,
  $P_{11} = -3.498$, $P_{02} = 2.306$.}. From this stage on, by using
the $u$-band magnitudes of stars in APASS derived as a function of the
$B$, $g$, $r$, and $i$ photometry, we proceed to derive and apply the
$u$-band correction maps as in $gri$ bands.

To further verify the validity of the calibration obtained with the
strategy delineated above, especially for the more elaborate $u$-band,
we matched and compared our photometry to the SkyMapper data \citep[SM
  hereafter;][]{wolf18,onken19}.  The SDSS photometric systems of
APASS and SM are not equivalent, the $u$ and $g$ bands in particular
show differences up to $0.5$ mag in the two systems
\citep{wolf18}. However, within the color interval $\lvert
g{-}i\rvert\leq 1$ mag, the SM to SDSS difference for $uri$-bands is
$\lsim0.1$ mag, while it is a factor of $\sim4$ larger in $g$-band
\citep[][see their Figure 17 and sections 2.2, 5.4]{wolf18}. Hence, as
a further consistency check, we compare our VST re-calibrated
photometry to SM data, within the color interval $\lvert
g{-}i\rvert\leq 1$ mag.

Over the entire FDS area covered with $ugri$ observations, we found
$\sim46,500$ sources in common with SM.  After identifying bright and
isolated stars, and with the given prescriptions on \gi color
selection, the final sample contains $\sim4,600$ objects ($\sim220$
per FDS field).

Table \ref{tab_vstsm} reports the median magnitude offsets between the
FDS and SM photometry for the matched sources, together with the $rms$
derived from the median absolute deviation\footnote{The median
  absolute deviation, MAD, defined as $\mathrm{MAD}=\mathrm{median}|
  {X}_{i}-\mathrm{median}(X)|$, is a robust indicator of the $rms$,
  which cleans the $rms$ from the spurious presence of few outliers in
  the sample. For a Gaussian distribution the standard deviation is
  $rms\sim1.48\times MAD$.}.  With the only not unexpected exception
of the $g$ band we find good agreement between the $u$, $r$ and $i$
photometry, with magnitude offsets better than 0.02 mag in $r$ and $i$
bands and of $\sim0.05$ mag in $u$; the $rms_{MAD}$ is $\sim0.03$ in
$gri$ and about twice larger in $u$-band.

%
\begin{table}
\caption{FDS magnitudes compared with APASS and SM}             
\label{tab_vstsm}      
\centering                          
\begin{tabular}{c c c c}        
\hline\hline                 
Filter & $rms_{VST-APASS}$ & $\Delta mag_{FDS-SM}$ & $rms_{VST-SM}$ \\    
\hline                        
   $u$ & 0.13 & $-0.054$ & 0.066    \\      
   $g$ & 0.03 & $0.149$  & 0.031   \\
   $r$ & 0.05 & $-0.015$ & 0.028   \\
   $i$ & 0.07 & $-0.003$ & 0.025   \\
\hline                                   
\end{tabular}
\end{table}

\begin{figure*}
   \centering
   \includegraphics[width=\textwidth]{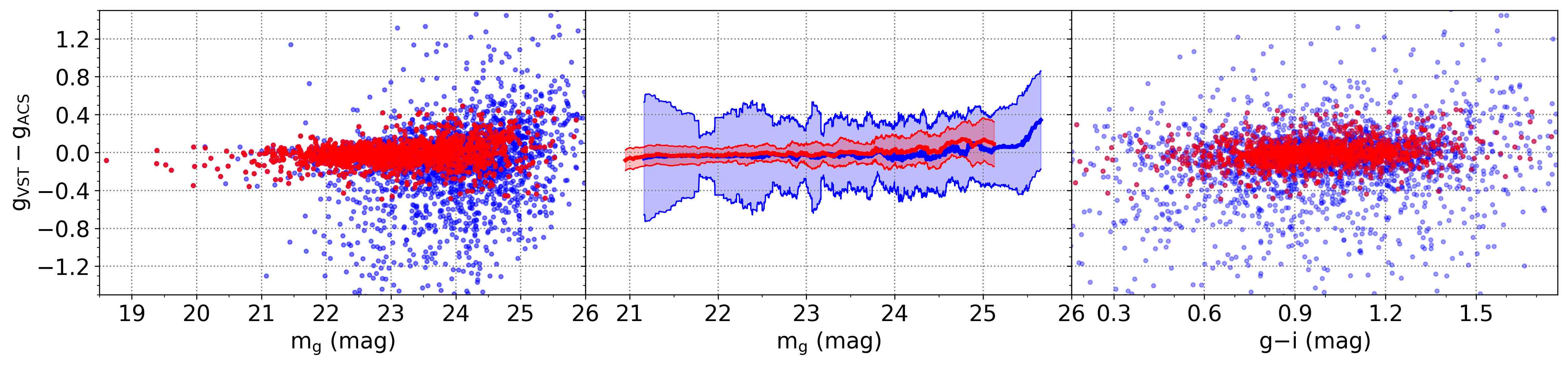}
      \caption{Left panel: $g$-band magnitudes from FDS compared with
        magnitudes of GC candidates from the ACSFCS. Blue symbols show
        the full matched set, red symbols identify compact sources in
        our reference catalog (see text).  Middle panel: as left
        panel, but running averages are shown, with bin size of 100/50
        objects for the blue/red symbols, respectively.  Right panel:
        as left panel, but versus \gi color.}
         \label{literature}
   \end{figure*}

For an independent check of the $g$-band photometry, we used the data
from the HST/ACS Fornax Cluster Survey
\citep[ACSFCS, ][]{jordan07,jordan15}. In Figure \ref{literature} we
report a comparison of our and ACSFCS $g$-band magnitudes.  We matched
the $\sim6.300$ GC candidates from the ACSFCS with the FDSex $gri$
catalog, to avoid the worse completeness limit of the $u$-band in the
$ugri$ catalogs. Adopting a matching radius of $1\farcs0$, a total of
3750 sources are found in common to both catalogs. The completeness of
the matching is $\sim90\%$ or higher at bright magnitudes
($m_g\leq23$), decreases to $\sim80\%$ for $m_g\leq24$, and is lower
than $\sim70\%$ for $m_g\leq25$. Hence, the completeness of the $gri$
catalog drops quickly below $m_g\sim24.5$ (mag), which corresponds to
$\sim$0.5 mag fainter than the turn over magnitude (TOM) of the GC
luminosity function (GCLF) for galaxies in Fornax \citep{villegas10}.

The left panel of Figure \ref{literature} shows the VST to ACSFCS
$g$-band magnitude difference versus $m_g$ (blue dots in the figure).
From the matched catalog, we selected a reference GC sample (see next
section), marked as red dots in the figure.  The running mean
difference for both the full matched sample and the reference sample
are shown in the middle panel, adopting window bin size 100/50 for the
full/best sample, respectively. Finally, the right panel of the
diagram shows the same quantities as in the left one, but versus the
\gi color. In all cases shown, the difference is consistent with zero
-- $\Delta g (FDS{-}ACSFCS)={-}0.03\pm0.12$ for the full sample of
3750 matched sources; $\Delta g (FDS{-}ACSFCS)={-}0.01\pm0.07$ for the
1455 sources in the reference catalog -- with no evidence of
significant residual trends.

\subsection{Morphometry}
\label{morphometry}

As already anticipated in section \S \ref{a-stacks}: with morphometry
we mean the measurement of all characteristics related to the shape of
the source, our reference frames for the morphological
characterization of sources are the multi-band $a$-stacks derived from
$gri$ exposures with best seeing.

We put particular emphasis on deriving quantities useful for
separating between point-like and extended sources, and identified a
number of useful features: FWHM, CLASS\_STAR, flux radius and
elongation (major-to-minor axis ratio) derived with SExtractor,
and the sharpness parameter derived from DAOphot.

For each source detected, we also measured the magnitude concentration
index, described in \citet{peng11}, defined as the difference in
magnitude measured at two different radial apertures. After various
tests we adopted as reference the concentration index derived from the
$a$-stacks aperture magnitudes at 4 and 6 pixels:
$CI=mag_{4~pix}-mag_{6~pix}$. For point-like sources, after applying
the aperture correction to the PSF magnitudes of isolated stars at
both radii, $CI$ should be statistically consistent with zero.  The
concentration index is constant for point-like objects, while extended
sources have variable $CI$ larger than zero.

\begin{figure*}
   \centering
   \includegraphics[width=18cm]{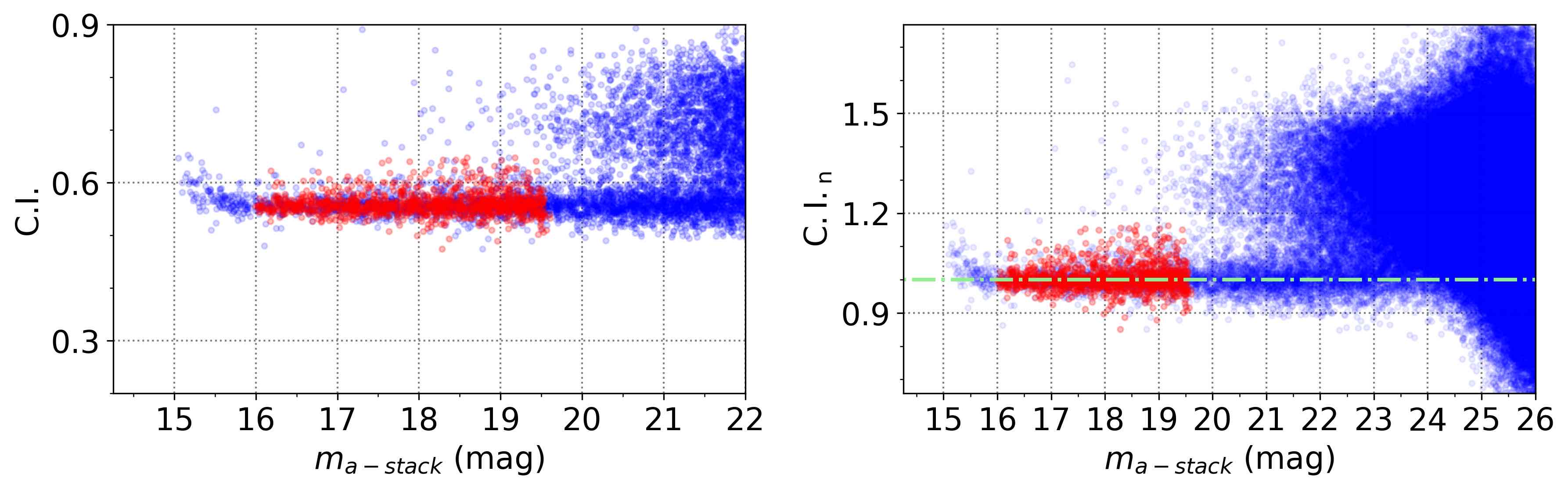}
      \caption{Normalization procedure for the CI, data for the field
        FDS\#13 are shown. Left panel: concentration index
        $CI=mag_{4~pix}-mag_{6~pix}$ versus the uncalibrated $a$-stack
        magnitude. For sake of clarity, only the brightest magnitude
        range is shown. Blue dots refer to the full sample, red
        symbols to candidate compact sources used to derive the median
        $CI$ factor for normalization. Right panel: as left panel, but
        over a larger magnitude range, and after normalization to the
        median $CI$ of bright point-like sources (red
        dots). Point-like sources candidates are aligned along the
        sequence parallel to the $x$-axis, around $CI{_n}\sim1$ (green
        dot-dashed line).}
         \label{cin}
   \end{figure*}

Because the $a$ is not a real photometric band, and because of the
field-to-field variations, for simplicity we decided to normalize the
CI index to 1, rather than to zero\footnote{The normalization to zero
  is the expected CI value for point-like sources after the proper
  aperture correction is applied to all sources. In our case, because
  the $a$-stacks are not in a real passband, and each FDS pointing has
  a different composition of good seeing $g$, $r$ and $i$ single
  exposures, we chose to avoid the aperture corrected normalization to
  zero.}. The normalization was derived as follows: for each field we
first estimated the CI from the magnitude difference within the two
chosen apertures (so no aperture correction is applied), then derived
the median CI of candidate point-like isolated and bright
sources. Finally, the CI of the full sample was normalized to the
median CI, such that compact sources should, by construction, be
characterized by normalized CI values, $CI_n$, of $\sim1$. Figure
\ref{cin} shows the procedure described, for sources in the field
FDS\#13: as expected compact sources (here selected using the
morphological parameters from SExtractor) occupy a flat sequence of
constant $CI$ (left panel), normalized to one in the right panel of
the figure.

\subsection{Final catalog and data quality}

The DAOphot and SExtractor catalogs of sources in the FDS fields are
finally combined in one single catalog (the same is done,
independently, for the FDSex regions). The final catalog contains:
$i)$ source identification adopting the IAU naming rules\footnote{See
  \url{https://www.iau.org/public/themes/naming/}.}, and position from
the $a$-stacks; $ii)$ the calibrated AB magnitudes from PSF photometry
derived with DAOphot in all available bands; $iii)$ the uncorrected
aperture and Kron-like magnitudes from SExtractor; $iv)$ the
morphometric parameters for $a$-stacks (FWHM, CLASS\_STAR, flux
radius, elongation and sharpness) and for all other available bands.

The FDS catalog provides data based on the 21 FDS field in the
$ugri$-bands, and for the $a$-stacks; a second $gri$-bands catalog for
the full FDSex area is also generated.

In the catalogs, we include the extinction correction term, assuming
the Galactic extinction values from the \citet{sf11} recalibration of
the \citet{sfd98} infrared based dust maps.

\begin{figure*}
   \centering \includegraphics[trim={3.3cm 1cm 2cm 0.5cm},clip,width=18.5cm]{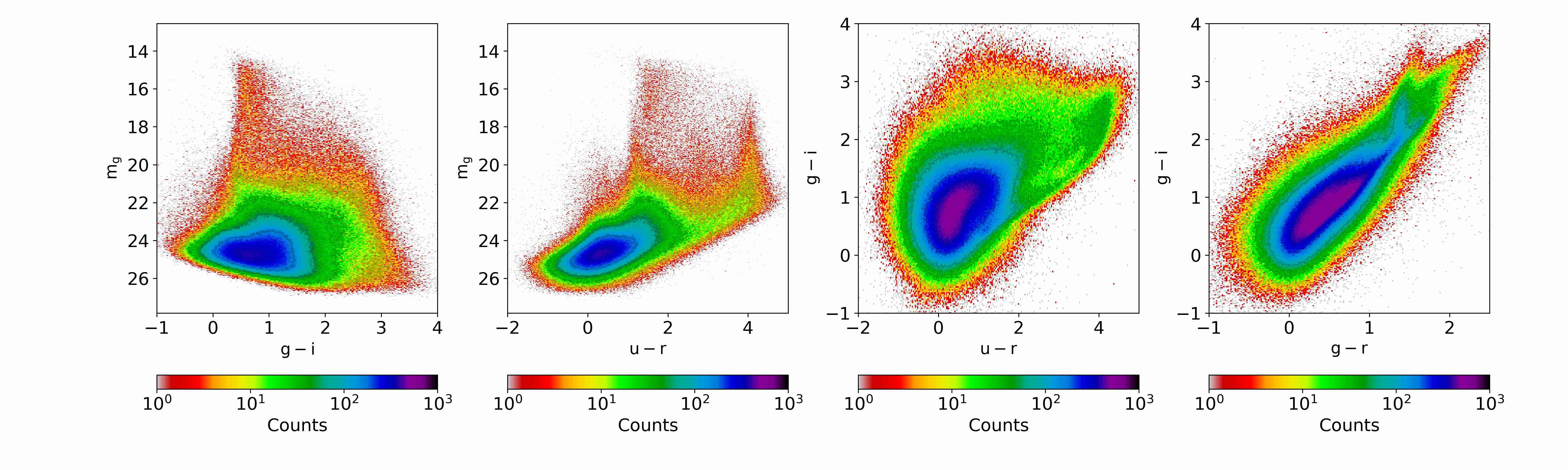}
      \caption{Hess color-magnitude and color-color diagrams of the
        full sample with $ugri$ photometry. Extinction corrected PSF
        magnitudes are used in all cases. FWHM is plotted in pixel
        units (the pixels scale is 0.2 arcsec/pixel).}
         \label{hess_all}
   \end{figure*}

Figure \ref{hess_all} shows a selection of extinction corrected color
magnitude and color-color diagrams for the full sample of sources in
the FDS catalog.

As overall photometric quality assessment we used the principal
colors, described in \citet{ivezic04}.  Principal colors are linear
combinations of the SDSS colors of stars. We adopted the coefficients
and selection parameters given in Tables 1-3 of \citet{ivezic04}. The
colors are combined to obtain a new color perpendicular to the stellar
locus. Assuming the position of the locus to be fixed, the value of
the principal colors is then an internal measure of the absolute
photometric calibration of the data.  Table \ref{tab_iq} provides the
median and $rms$ width of three principal colors, $P2(s)$, $P2(w)$ and
$P2(x)$ for each FDS field; the median P2 values over the full
  set of fields is $<0.02$ with $rms\lsim0.03$. The $P_2(s)$ depends
on the $u$-band photometry, and cannot be determined over the FDSex
fields.  The overall $\langle P_2\rangle$ and $\sigma[P_2]$ values,
and the values for each field, are consistent with the same value
reported by \citet{ivezic04} for SDSS photometry.

Finally, we obtain the limiting magnitudes reported in Table
\ref{tab_iq} for all fields and bands, derived as $5\sigma$ magnitude
integrated over the PSF, determined from the median S/N estimated as
$\Delta m_{PSF}^{-1}$. The median $g$-band limiting magnitude is
$g_{lim}\sim25.4\pm0.2$ mag; we note that the faintest GCs matched
with the ACSFCS reach $m_g\sim25.6$ mag, which increases to
$m_g\sim25.2$ mag for the sources in the reference catalog.

All catalogs are available via a dedicated web-interface of the FDS
team\footnote{\url{fdscat.oa-abruzzo.inaf.it}}, and
are made available through the CDS. An extract of the data for the
$\sim$1.7 million $ugri$ matched sources in the FDS catalog is
reported in Table \ref{tab_ugri} (an extract for the $\sim$3.1 million
sources in the FDSex $gri$ catalog is given in Table \ref{tab_gri}).
%

\begin{sidewaystable}
\tiny
\caption{An extract of the FDS $ugri$ catalog}            
\label{tab_ugri}   
\centering                          
\begin{tabular}{ccccccccccccccc}
\hline\hline                 
          ID              &  R.A. (J2000)& Dec. (J2000)   & $m_u$                     &     $m_g$                 &   $m_r$                 &    $m_i$                  & Star/Gal.   &    $CI_n$ &  F.R.    &   FWHM      &  Elong.&  Sharp.   & $E(B{-}V)$ & Field \\
                          &   (deg)      &   (deg)        & (mag)                     &     (mag)                 &    (mag)                &     (mag)                 &            &           & (pixel) & (pixel)    &            &             &            & (\#) \\
        (1)               &   (2)        &   (3)          & (4)                       &     (5)                   &    (6)                  &     (7)                   &    (8)     &    (9)    &   (10)   & (11)        &  (12)      &   (13)      &  (14)      & (15) \\
\hline                        
FDSJ033744.66-345442.84   &   54.436077  &   -34.911900   &    21.550  $\pm$  0.051   &    20.013 $\pm$  0.053    &   19.172 $\pm$  0.066   &    18.507 $\pm$  0.059   &     0.029  &    1.449  &   12.57   &   13.84    &    1.260   &      4.559   &     0.011  &   10 \\
FDSJ034014.64-345451.44   &   55.061020  &   -34.914288   &    21.675  $\pm$  0.042   &    20.368 $\pm$  0.039    &   19.509 $\pm$  0.048   &    18.865 $\pm$  0.037   &     0.029  &    1.266  &    7.37   &   12.47    &    1.670   &      3.240   &     0.012  &   10 \\
FDSJ034028.79-345508.98   &   55.119957  &   -34.919163   &    18.585  $\pm$  0.006   &    15.914 $\pm$  0.005    &   14.790 $\pm$  0.010   &    14.217 $\pm$  0.006   &     0.996  &    1.040  &    3.25   &    4.56    &    1.010   &      0.872   &     0.012  &   10 \\
FDSJ034055.30-345511.56   &   55.230427  &   -34.919880   &    24.431  $\pm$  0.192   &    24.853 $\pm$  0.124    &   24.014 $\pm$  0.124   &    23.565 $\pm$  0.154   &     0.830  &    1.243  &    3.44   &    7.99    &    1.570   &      0.920   &     0.011  &   10 \\
FDSJ034044.55-345510.26   &   55.185608  &   -34.919518   &    25.478  $\pm$  0.511   &    24.972 $\pm$  0.116    &   24.841 $\pm$  0.228   &    25.056 $\pm$  0.507   &     0.512  &    1.304  &    2.26   &    3.45    &    1.700   &      1.218   &     0.011  &   10 \\
FDSJ034036.60-345510.96   &   55.152508  &   -34.919712   &    24.901  $\pm$  0.306   &    25.313 $\pm$  0.171    &   25.395 $\pm$  0.427   &    24.588 $\pm$  0.348   &     0.562  &    0.659  &    1.67   &    2.00    &    1.340   &     -5.609   &     0.011  &   10 \\
FDSJ034055.72-345510.38   &   55.232162  &   -34.919552   &    25.046  $\pm$  0.358   &    24.225 $\pm$  0.058    &   23.586 $\pm$  0.074   &    22.863 $\pm$  0.089   &     0.875  &    1.020  &    2.59   &    8.38    &    1.130   &      1.078   &     0.011  &   10 \\
FDSJ034107.11-345506.88   &   55.279636  &   -34.918579   &    24.675  $\pm$  0.284   &    25.149 $\pm$  0.178    &   24.641 $\pm$  0.258   &    24.698 $\pm$  0.487   &     0.699  &    0.737  &    1.68   &    4.38    &    1.640   &     -6.010   &     0.011  &   10 \\
FDSJ034015.36-345510.38   &   55.063984  &   -34.919552   &    25.319  $\pm$  0.471   &    26.016 $\pm$  0.316    &   24.720 $\pm$  0.172   &    24.554 $\pm$  0.457   &     0.430  &    1.082  &    2.10   &    3.79    &    1.200   &     -0.228   &     0.012  &   10 \\
FDSJ033954.02-345511.46   &   54.975067  &   -34.919849   &    24.812  $\pm$  0.326   &    24.716 $\pm$  0.094    &   24.663 $\pm$  0.168   &    23.768 $\pm$  0.167   &     0.643  &    1.398  &    3.71   &    8.84    &    2.980   &     -1.337   &     0.012  &   10 \\
FDSJ034037.01-345511.86   &   55.154224  &   -34.919960   &    24.901  $\pm$  0.285   &    23.111 $\pm$  0.026    &   22.269 $\pm$  0.027   &    21.868 $\pm$  0.038   &     0.980  &    0.956  &    2.69   &    4.55    &    1.080   &     -0.396   &     0.011  &   10 \\
FDSJ033612.00-345454.11   &   54.050011  &   -34.915031   &    21.314  $\pm$  0.059   &    20.323 $\pm$  0.066    &   19.940 $\pm$  0.081   &    19.444 $\pm$  0.072   &     0.029  &    1.543  &    9.89   &   20.07    &    1.240   &      5.941   &     0.013  &   10 \\
\hline                                   
\end{tabular}
\tablefoot{Columns list:(1) FDS ID; (2) Right Ascension; (3)
  Declination; (4-7) $ugri$-band magnitude with error; (8) Star/Galaxy
  classifier, CLASS\_STAR, from SExtractor; (9) normalized
  concentration index; (10) Flux Radius, from SExtractor in
  pixels (pixel scale is 0.2 arcsec/pixel); (11) FWHM in pixels; (12) Elongation, major-to-minor
  axis ratio; (13) DAOphot sharpness parameter; (14) Reddening from
  \citet{sf11} ; (15) FDS field pointing ID. All morphological
  quantities from Cols. (8-13) are derived from the $a$-stacks.  The
  full table is available in electronic form at the CDS, and at the
  web-pages of the project, at:
  \url{fdscat.oa-abruzzo.inaf.it}.}
\end{sidewaystable}

\begin{sidewaystable}
\tiny
\caption{An extract of the FDSex $gri$ catalog}            
\label{tab_gri}   
\centering                          
\begin{tabular}{cccccccccccccc}
\hline\hline                 
          ID              &  R.A. (J2000)& Dec. (J2000)   &     $m_g$                 &   $m_r$                    &    $m_i$                & Star/Gal.   &    $CI_n$ &  F.R.    &   FWHM      &  Elong.&  Sharp.   & $E(B{-}V)$ & Field \\
                          &   (deg)      &   (deg)        & (mag)                     &     (mag)                 &    (mag)                &             &           & (pixel)   & (pixel)    &            &             &            & (\#) \\
        (1)               &   (2)        &   (3)          & (4)                       &     (5)                   &    (6)                  &     (7)                   &    (8)     &    (9)    &   (10)   & (11)        &  (12)      &   (13)      &  (14) \\
\hline                        
FDSJ033332.60-374029.36  &    53.385815 &  -37.674820  &   25.154  $\pm$   0.127 &    25.010  $\pm$   0.185  &   24.028  $\pm$   0.175  &    0.566 &   1.169  &  3.21  &  10.39 &    1.130   &    1.797  &   0.011 &  18\\
FDSJ033343.98-374044.61  &    53.433262 &  -37.679058  &   20.012  $\pm$   0.010 &    19.071  $\pm$   0.006  &   18.451  $\pm$   0.005  &    0.956 &   1.066  &  3.16  &   4.83 &    1.020   &    0.669  &   0.011 &  18\\
FDSJ033138.97-374012.90  &    52.912369 &  -37.670250  &   24.498  $\pm$   0.077 &    23.868  $\pm$   0.093  &   23.171  $\pm$   0.140  &    0.015 &   1.228  &  3.55  &   9.99 &    1.490   &    2.733  &   0.016 &  18\\
FDSJ033539.47-374048.47  &    53.914478 &  -37.680130  &   23.700  $\pm$   0.073 &    22.854  $\pm$   0.054  &   22.144  $\pm$   0.042  &    0.018 &   1.428  &  4.72  &  11.84 &    1.340   &    3.798  &   0.015 &  18\\
FDSJ033421.67-374037.64  &    53.590298 &  -37.677124  &   24.604  $\pm$   0.093 &    24.398  $\pm$   0.112  &   23.846  $\pm$   0.190  &    0.125 &   1.307  &  3.23  &   9.30 &    1.370   &    2.152  &   0.013 &  18\\
FDSJ033258.65-374024.76  &    53.244354 &  -37.673546  &   24.396  $\pm$   0.074 &    24.348  $\pm$   0.105  &   24.330  $\pm$   0.232  &    0.457 &   1.111  &  2.50  &   6.21 &    1.120   &    0.839  &   0.015 &  18\\
FDSJ033118.49-374009.10  &    52.827042 &  -37.669193  &   24.306  $\pm$   0.083 &    23.514  $\pm$   0.075  &   22.892  $\pm$   0.088  &    0.012 &   1.374  &  4.59  &  15.56 &    1.330   &    4.176  &   0.016 &  18\\
FDSJ033216.89-374016.07  &    53.070377 &  -37.671131  &   25.622  $\pm$   0.183 &    24.934  $\pm$   0.170  &   24.127  $\pm$   0.226  &    0.474 &   1.368  &  2.91  &   5.70 &    1.310   &    2.713  &   0.016 &  18\\
FDSJ033258.88-374023.19  &    53.245346 &  -37.673107  &   25.225  $\pm$   0.144 &    25.426  $\pm$   0.294  &   25.130  $\pm$   0.500  &    0.513 &   1.013  &  1.86  &   4.39 &    2.190   &   -1.384  &   0.015 &  18\\
FDSJ033211.86-374023.13  &    53.049416 &  -37.673092  &   24.437  $\pm$   0.096 &    22.781  $\pm$   0.057  &   21.589  $\pm$   0.063  &    0.001 &   1.441  &  6.47  &  19.87 &    1.390   &    4.409  &   0.015 &  18\\
FDSJ033157.13-374013.28  &    52.988060 &  -37.670353  &   25.117  $\pm$   0.144 &    24.759  $\pm$   0.180  &   24.338  $\pm$   0.269  &    0.559 &   1.473  &  4.00  &   8.01 &    1.700   &    3.462  &   0.015 &  18\\
FDSJ033107.58-374005.39  &    52.781567 &  -37.668163  &   24.658  $\pm$   0.083 &    23.182  $\pm$   0.051  &   22.461  $\pm$   0.063  &    0.798 &   1.112  &  2.92  &   7.75 &    1.330   &    1.349  &   0.016 &  18\\
\hline                                   
\end{tabular}
\tablefoot{Columns list:(1) FDS ID; (2) Right Ascension; (3)
  Declination; (4-6) $gri$-band magnitude with error; (7) Star/Galaxy
  classifier, CLASS\_STAR, from SExtractor; (8) normalized
  concentration index; (9) Flux Radius, from SExtractor in pixel (piscel scale is 0.2 arcsec/pixel);
  (10) FWHM in pixel; (11) Elongation, major-to-minor axis ratio;
  (12) DAOphot sharpness parameter; (13) Reddening from \citet{sf11} ;
  (14) FDS field pointing ID. All morphological quantities from Cols.
  (7-12) are derived from the $a$-stacks.  The full table is available
  in electronic form at the CDS, and at the web-pages of the project,
  at:
  \url{fdscat.oa-abruzzo.inaf.it}.}
\end{sidewaystable}

   \begin{figure*}
   \centering
   \includegraphics[width=18cm]{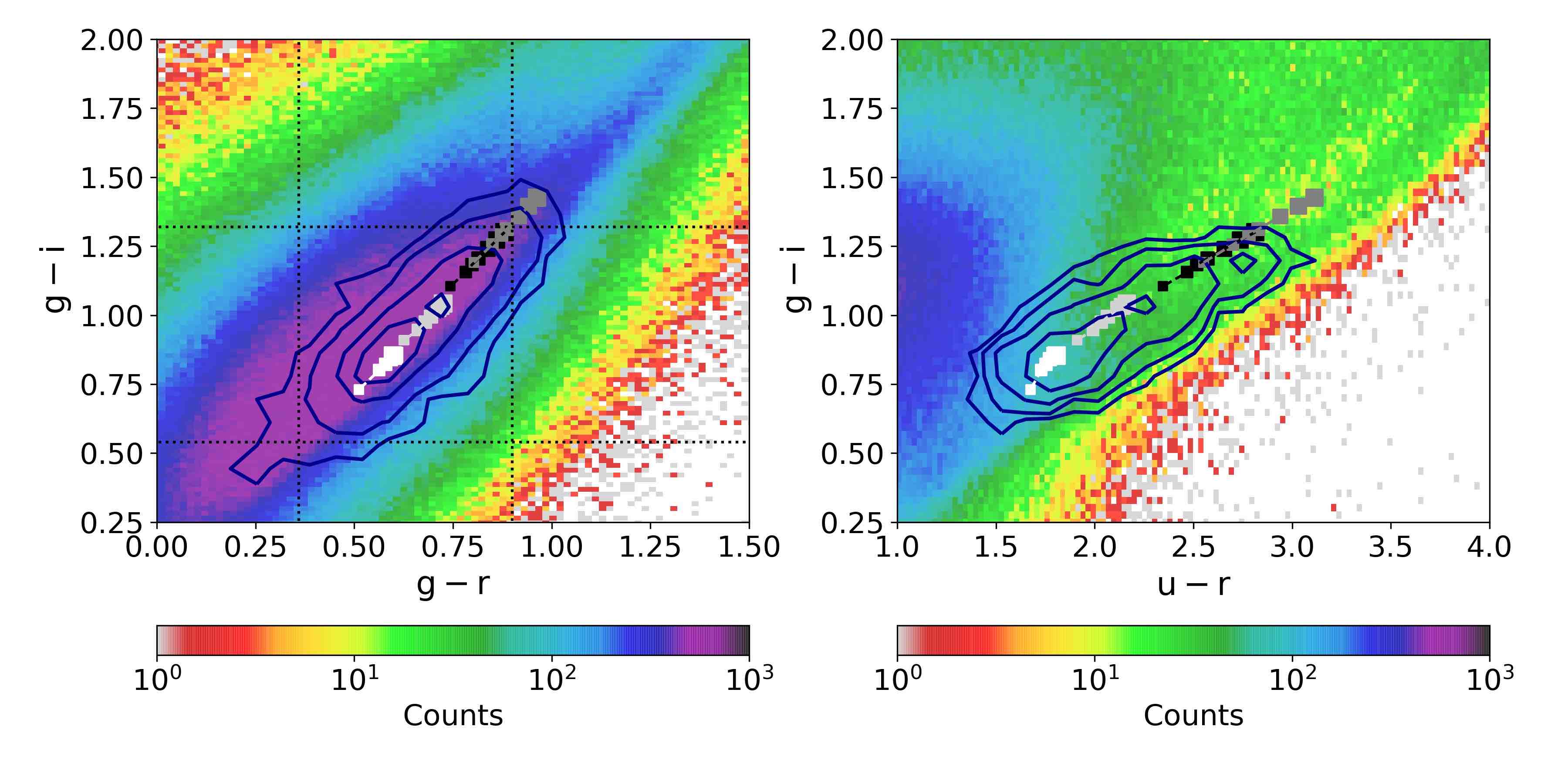}
   \includegraphics[width=18cm]{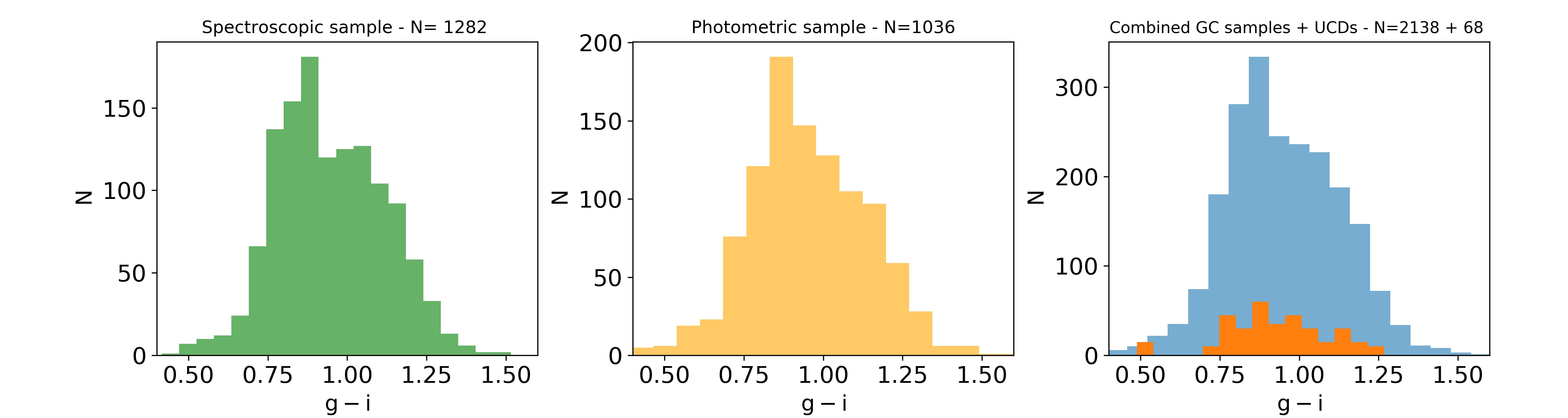}
   \includegraphics[width=18cm]{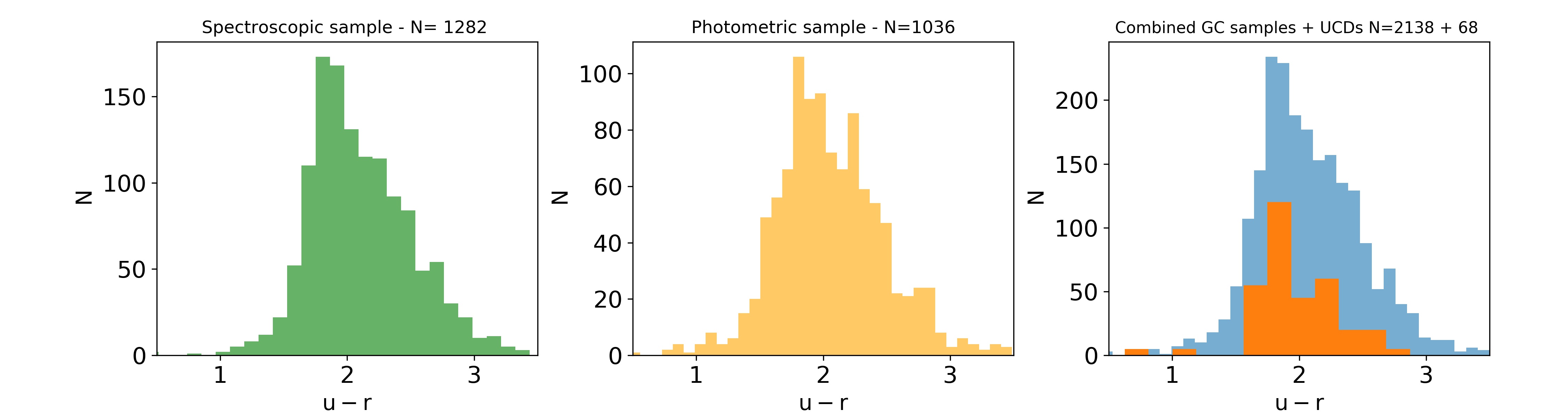}
      \caption{Upper panels: Color-color Hess diagrams for the sample
        of sources with $ugri$ photometry, over the color interval
        expected for GCs and UCDs. The dark-blue lines show the linear
        spaced contour levels of sources in the master GC catalog.
        Filled squares show the integrated colors from the SPoT
        stellar population synthesis code. White, light-gray, black
        and dark-gray symbols indicate metallicity \feh=[-1.3, -0.7,
          0.0, 0.4] respectively; symbols size scales with increasing
        model age, ranging between 4 and 14 Gyr, with 2 Gyr step. Same
        metallicity models are connected with dashed lines.  In the
        left color-color panel we also draw with dotted lines the
        color intervals of GCs assuming $\pm3{-}rms_{MAD}$ with
        respect to the median values in Table \ref{tab_medmad}. Middle
        and lower panels: \gi and \ur color histograms, respectively,
        for the master spectroscopic (left panel, green histogram),
        photometric (middle, yellow), and combined (right, blue) GC
        catalogs.  In the third histograms the data of UCDs are also
        shown (orange), expanded by a factor of five for sake of
        clarity. Only sources brighter than $m_g=23.5$ mag are
        considered.}
         \label{hess_histo}
   \end{figure*}

\section{A preliminary map of GCs and UCD galaxies over the FDS area}
\label{maps}

One of the goals of the FDS survey is to map the distribution of GCs
and UCDs in Fornax out to the virial radius. In the following
sections, and --more in details-- in a forthcoming dedicated paper
(Cantiello et al., 2020, in prep.) we analyze and discuss the
cluster-wide properties of these two classes of compact stellar
systems, with more emphasis on GCs.

Unambiguously identifying GCs from purely optical photometry is
unfeasible. In \citet{cantiello18n253} we showed that also
spectroscopic samples might be affected by non negligible
contamination. \citet{munoz14} demonstrated that optical data
including the $u$ band, combined with $K$-band near-IR data can
dramatically reduce the contamination by fore/background sources.

Lacking a publicly available deep near-IR survey covering the FDS
area, we proceed as already done in our previous works on GCs from the
VEGAS and FDS surveys
  \citep{cantiello15,cantiello16,cantiello18vegas,dabrusco16}.

Briefly, we identify a master catalog of GCs, and UCDs, and use the
main properties of confirmed sources to constrain the mean loci
of several photometric (magnitudes, colors, etc.) and morphometric
($CI_n$, galaxy/star classification, etc.) indicators.

In the following section we discuss the procedures adopted for
identifying the loci of GCs using several parameters.

\subsection{GCs \& UCDs Master Catalogs}

We define a master catalog of GCs and UCDs taking as reference
literature spectroscopic and photometric studies, adopting $M_g=-10.5$
mag as GC/UCD separation criteria, corresponding to $M_V\sim-11$ mag
($\sim10^7 M_{\odot}$), and to an apparent $m_g=21$ magnitude at the
adopted distance to Fornax \citep[e.g.][]{mieske04,hilker07}.  We
collect photometric data from the already mentioned ACSFCS survey
\citep{jordan07, jordan15}. The advantage of ACS with respect to other
imagers, is the very high resolution allowed by the space-based
observations. At the distance of Fornax, GCs observed with the ACS
camera appear as partially resolved sources, so their physical size
can be estimated and used as a further parameter to reliably separate
them from foreground stars and background galaxies.  From the ACSFCS
GC sample, we selected only GC candidates with high probability
$p_{GC}$ of being a GC \citep[$p_{GC}\geq0.75$, derived according to a
  maximum-likelihood estimate,][]{jordan09}.

The spectroscopic sample is a combination of \citet{pota18} and
\citet{schuberth10} datasets. By matching the spectroscopic and
photometric catalogs --cleaned up by the common sources-- with our FDS
$ugri$ catalog, we obtained a list of $\sim$3.250 GCs. 

%

We complete our master catalogs of reference compact stellar systems
with 68 bright sources in Fornax, confirmed UCD compiled from the
available spectroscopic and photometric literature for this class of
objects in Fornax.

The GC and UCD master catalogs are given in the Table
\ref{tab_GCmaster} and \ref{tab_UCDmaster}.

The upper panels in Figure \ref{hess_histo} shows the same color-color
diagrams as in Figure \ref{hess_all} with a zoom over the color-color
region of GCs and UCDs. The contour levels of sources from the master
catalog are reported with thick dark-blue lines (we adopt linear
  spacing for contour levels). In the figure we also report the SPoT
simple stellar population models
\citep{brocato99,cantiello03,raimondo05}, for an age range of 4-14 Gyr
and metallicity \feh=-1.3 to 0.4 dex. The consistency between the
empirical loci of GCs and stellar population models for the typical
age and metallicity ranges of GCs, provides further independent
support to the reliability of the calibration approach adopted.
  In the \ur-\gi plane, the most metal-rich old stellar population
  models do not match with the observed GC distribution. One possible
  explanation is the combination of two effects: the small number of
  observed old GCs with such high metallicity (age$\geq10$ Gyr,
  $\feh=0.4$, more than twice solar metallicity), and, consequently,
  the uncertainties of stellar population models is this regime.

The middle and lower panels of the figure also show the \gi and \ur
color histograms for the photometric, spectroscopic and combined
samples, for sources brighter than $m_g=23.5$ mag.  The asymmetric
appearance of the color distribution is a consequence of the
well-known color bimodality of GC systems in some filters
\citep{ashman92,yoon06,blake10a,usher12,cantiello14}, here smoothed as
the GC sample is a combination of GCs around $\sim30$ galaxies in
Fornax, each one with different morphological types and magnitudes,
hence with different properties in terms of GCs color peaks
\citep{peng06}.

\begin{sidewaystable}
  \caption{Master catalog of GCs}            
  \tiny
\label{tab_GCmaster}   
\centering                          
\setlength{\tabcolsep}{2pt}
\begin{tabular}{ccccccccccccccccccc}
\hline\hline                 
         ID            &R.A. (J2000)& Dec. (J2000)&        $m_u$           & $m_g$                   &   $m_r$                  &    $m_i$                & Star/Gal. & $CI_n$ &  F.R. &   FWHM &  Elong.&  Sharp. & $E(B{-}V)$ & Field &FCC&$p_{GC}$&$r_h$& Source \\
                       &   (deg)    &   (deg)     & (mag)                  &     (mag)               &    (mag)                 &    (mag                &           &        &(pixel) &(pixel)&        &         &            & (\#)  &   &     &(arcsec)  &       \\
        (1)            &   (2)      &   (3)       & (4)                    &     (5)                 &    (6)                   &     (7)                &    (8)    &    (9) &   (10)& (11)   &  (12)  &   (13)  &  (14)      &(15)   &(16)&(17)&(18) & (19)  \\
\hline                        
FDS032625.46-354235.74 &  51.606079 &  -35.709927 &  24.917(0.449)&   23.694(0.059) &   22.952(0.047)  &  22.718(0.095) &   0.887 &  1.048 &  2.26  &   5.64  &  1.14 &  0.565 &   0.01  &  20  &   47  &  1.0  &  0.243 &  P \\             
FDS032626.39-354229.81 &  51.609943 &  -35.708279 &  24.595(0.336)&   22.749(0.022) &   22.013(0.025)  &  21.688(0.037) &   0.981 &  0.998 &  2.61  &   4.32  &  1.12 &  0.141 &   0.01  &  20  &   47  &  1.0  &  0.275 &  P \\            
FDS032627.14-354245.15 &  51.613079 &  -35.712543 &  24.001(0.213)&   22.418(0.02 ) &   21.774(0.019)  &  21.528(0.036) &   0.983 &  1.039 &  2.44  &   4.24  &  1.04 &  0.304 &   0.01  &  20  &   47  &  1.0  &  0.281 &  P \\            
FDS032627.18-354357.56 &  51.613262 &  -35.732655 &  23.739(0.139)&   22.258(0.017) &   21.39 (0.015)  &  21.144(0.024) &   0.962 &  1.017 &  2.6   &   4.25  &  1.06 &  0.387 &   0.01  &  20  &   47  &  1.0  &  0.313 &  P \\             
FDS032627.23-354125.68 &  51.613449 &  -35.690468 &  24.354(0.238)&   23.268(0.039) &   22.535(0.031)  &  22.367(0.065) &   0.982 &  1.101 &  2.58  &   4.75  &  1.06 &  0.665 &   0.01  &  20  &   47  &  1.0  &  0.317 &  P \\             
FDS032627.27-354237.98 &  51.613628 &  -35.710552 &  24.623(0.332)&   24.115(0.083) &   23.461(0.074)  &  23.201(0.124) &   0.746 &  1.056 &  2.7   &   6.41  &  1.17 &  1.172 &   0.01  &  20  &   47  &  0.95 &  0.47  &  P \\             
FDS032627.38-354224.34 &  51.614079 &  -35.70676  &  24.743(0.404)&   23.363(0.039) &   22.779(0.039)  &  22.535(0.078) &   0.979 &  1.029 &  2.63  &   5.02  &  1.13 &  0.603 &   0.01  &  20  &   47  &  0.98 &  0.529 &  P \\             
FDS032627.66-354441.02 &  51.615265 &  -35.744728 &  23.791(0.134)&   22.496(0.02 ) &   21.671(0.018)  &  21.459(0.029) &   0.982 &  0.987 &  2.51  &   4.14  &  1.05 &  0.082 &   0.01  &  20  &   47  &  1.0  &  0.283 &  P \\             
FDS032628.10-354356.31 &  51.617069 &  -35.732307 &  24.739(0.291)&   23.323(0.039) &   22.408(0.032)  &  22.218(0.059) &   0.977 &  1.036 &  2.35  &   4.55  &  1.05 &  0.235 &   0.01  &  20  &   47  &  1.0  &  0.356 &  P \\             
FDS032628.17-354359.17 &  51.617355 &  -35.733105 &  25.015(0.486)&   23.795(0.058) &   23.191(0.055)  &  22.928(0.086) &   0.177 &  1.637 &  5.3   &   16.93 &  1.43 &  1.453 &   0.01  &  20  &   47  &  0.98 &  0.37  &  P \\             
FDS032628.20-354425.43 &  51.617496 &  -35.740398 &  24.222(0.194)&   24.278(0.081) &   23.706(0.095)  &  23.046(0.113) &   0.656 &  1.115 &  2.65  &   7.61  &  1.21 &  0.613 &   0.01  &  20  &   47  &  0.97 &  0.218 &  P \\             
FDS032628.34-354341.85 &  51.618095 &  -35.728291 &  25.107(0.395)&   24.443(0.102) &   23.598(0.075)  &  23.546(0.185) &   0.8   &  1.048 &  2.49  &   5.77  &  1.14 &  0.581 &   0.01  &  20  &   47  &  0.99 &  0.294 &  P \\             
\nodata                &            &             &               &                 &                      &                      &         &        &        &         &       &        &         &      &       &       &        &    \\             
FDS033633.14-345643.64 &  54.138096 &  -34.945457 &  23.595(0.142)&   22.007(0.013)  &  21.226(0.014)  &  20.954(0.016) &   0.921 &  1.007 &  2.76  &   4.29  &  1.08 &  0.341 &   0.015 &  11  & \nodata & \nodata &  \nodata & S \\
FDS033633.49-350248.19 &  54.139542 &  -35.046719 &  24.75 (0.418)&   22.906(0.031)  &  22.175(0.027)  &  21.732(0.045) &   0.984 &  1.013 &  2.5   &   4.39  &  1.03 &  0.27  &   0.014 &  11  & \nodata & \nodata &  \nodata & S  \\     
\nodata                &            &             &               &                &                   &                &         &        &        &         &       &        &         &      &       &       &        &    \\             
FDS033630.08-350013.69 &  54.125324 &  -35.003803 &  24.466(0.393)&  22.079(0.014 ) &   21.266(0.016)   & 20.862(0.021) &   0.926 &  1.054 &  2.81  &   4.75  &  1.1 &   0.486 &   0.015 &  11  &   167   &1.0      & 0.381   & S+P \\           
FDS033630.10-351753.79 &  54.125427 &  -35.298275 &  22.792(0.066)&  21.364(0.009 ) &   20.771(0.012)   & 20.453(0.019) &   0.911 &  1.084 &  3.3   &   5.18  &  1.53&   0.489 &   0.011 &  11  &   170   &1.0      & 0.468   & S+P \\         
\hline                                   
\end{tabular}
\tablefoot{Columns list:(1) FDS ID; (2) Right Ascension; (3)
  Declination; (4-7) $ugri$-band magnitude with error; (8) Star/Galaxy
  classifier, CLASS\_STAR, from SExtractor; (9) normalized
  concentration index; (10) Flux Radius, from SExtractor in arcsec;
  (11) FWHM in arcseconds; (12) Elongation, major-to-minor axis ratio;
  (13) DAOphot sharpness parameter; (14) Reddening from \citet{sf11} ;
  (15) FDS field pointing ID. All morphological quantities from
  Cols. (8-13) are derived from the $a$-stacks. (16-18) Fornax cluster
  catalog ID of the host galaxy, $p_{GC}$ likelihood, and median $g$
  and $z$ GC half light radius from \citet{jordan15}; (19) Source of the
  confirmed GC: ``S'' for spectroscopic confirmed GC --from
  \citet{pota18} or \citet{schuberth10}--, ``P'' for photometric
  confirmed GC from the ACSFCS dataset.  The full table is available
  in electronic form at the CDS.}
\end{sidewaystable}

\begin{sidewaystable}
  \caption{Master catalog of UCDs}            
  \tiny
\label{tab_UCDmaster}   
\centering                          
\setlength{\tabcolsep}{3pt}
\begin{tabular}{ccccccccccccccccc}
\hline\hline                 
         ID            &R.A. (J2000)& Dec. (J2000)&        $m_u$         & $m_g$            &   $m_r$           &    $m_i$       & Star/Gal. & $CI_n$ &  F.R. &   FWHM &  Elong.&  Sharp. & $E(B{-}V)$ & Field & $v_{hel}$& Source \\
                       &   (deg)    &   (deg)     & (mag)               &     (mag)         &    (mag)          &    (mag       &           &        &(pixel)&(pixel)&        &         &            & (\#)  & (km/s) &       \\
        (1)            &   (2)      &   (3)       & (4)                &     (5)            &    (6)            &     (7)       &    (8)    &    (9) &   (10)& (11)   &  (12)  &   (13)  &  (14)      &(15)   &(16)    &(17)   \\
\hline                        
FDS033854.05-353333.42 &  54.725212  & -35.559284  & 20.701(0.037) &   18.959 (0.042)  &  18.037(0.041) &   17.626(0.036) &   0.029  & 1.321  & 6.05  & 7.77  &   1.04  &  3.393  &   0.01  &  11   &   1517.0 (6.0) & F08\\
FDS033805.05-352409.33 &  54.521023  & -35.402592  & 21.056(0.015) &   19.308 (0.007)  &  18.568(0.009) &   18.238(0.007) &   0.959  & 1.024  & 2.99  & 4.52  &   1.13  &  0.559  &   0.012 &  11   &   1198.9 (6.1) & F08\\
FDS033935.92-352824.59 &  54.899654  & -35.473499  & 21.109(0.022) &   19.673 (0.027)  &  19.033(0.022) &   18.588(0.013) &   0.261  & 1.163  & 3.78  & 5.31  &   1.06  &  1.585  &   0.011 &  11   &   1878.0 (5.0) & B07\\
FDS033806.29-352858.72 &  54.526222  & -35.482979  & 21.447(0.022) &   19.814 (0.017)  &  19.086(0.022) &   18.681(0.014) &   0.799  & 1.137  & 3.59  & 5.39  &   1.26  &  1.335  &   0.011 &  11   &   1234.0 (5.0) & B07\\
FDS033703.22-353804.51 &  54.263435  & -35.634586  & 21.66 (0.025) &   19.895 (0.015)  &  19.12 (0.018) &   18.699(0.013) &   0.537  & 1.123  & 3.61  & 4.94  &   1.03  &  1.482  &   0.011 &  11   &   1561.0 (3.0) & F08\\
FDS033810.34-352405.79 &  54.543095  & -35.401608  & 21.653(0.019) &   19.987 (0.005)  &  19.261(0.005) &   18.972(0.005) &   0.979  & 0.985  & 2.71  & 4.18  &   1.09  &  0.131  &   0.012 &  11   &   1626.0 (10.0) & M08\\
FDS033952.54-350424.04 &  54.968903  & -35.073345  & 21.342(0.023) &   20.018 (0.019)  &  19.332(0.014) &   19.069(0.018) &   0.863  & 1.073  & 3.71  & 4.63  &   1.06  &  0.955  &   0.011 &  11   &   1236.0 (21.0) & F08\\
FDS033823.72-351349.49 &  54.59885   & -35.230415  & 21.515(0.02 ) &   20.13  (0.007)  &  19.579(0.01 ) &   19.308(0.013) &   0.831  & 1.117  & 3.12  & 4.74  &   1.02  &  1.129  &   0.012 &  11   &   1637.0 (14.0) & F08\\
FDS033743.56-352251.47 &  54.431484  & -35.380966  & 21.616(0.025) &   20.16  (0.007)  &  19.592(0.014) &   19.271(0.011) &   0.764  & 1.093  & 3.23  & 4.81  &   1.07  &  1.214  &   0.013 &  11   &   1420.0 (7.0) & F08\\
FDS033841.94-353313.03 &  54.674747  & -35.553619  & 21.944(0.025) &   20.194 (0.01 )  &  19.462(0.008) &   19.091(0.007) &   0.967  & 1.051  & 2.86  & 4.47  &   1.04  &  0.578  &   0.01  &  11   &   2024.0 (10.0) & M08\\
FDS033627.70-351413.84 &  54.115421  & -35.237179  & 22.214(0.046) &   20.198 (0.009)  &  19.36 (0.02 ) &   18.925(0.02 ) &   0.803  & 1.171  & 3.45  & 5.55  &   1.14  &  1.385  &   0.012 &  11   &   1386.0 (4.0) & B07\\
FDS033920.51-351914.25 &  54.835464  & -35.320625  & 21.916(0.032) &   20.253 (0.02 )  &  19.562(0.025) &   19.031(0.013) &   0.176  & 1.174  & 3.61  & 5.37  &   1.03  &  1.905  &   0.011 &  11   &   1462.0 (5.0) & F08\\
\hline                                   
\end{tabular}
\tablefoot{Columns list:(1) FDS ID; (2) Right Ascension; (3)
  Declination; (4-7) $ugri$-band magnitude with error; (8) Star/Galaxy
  classifier, CLASS\_STAR, from SExtractor; (9) normalized
  concentration index; (10) Flux Radius, from SExtractor in arcsec;
  (11) FWHM in arcseconds; (12) Elongation, major-to-minor axis ratio;
  (13) DAOphot sharpness parameter; (14) Reddening from \citet{sf11} ;
  (15) FDS field pointing ID. All morphological quantities from
  Cols. (8-13) are derived from the $a$-stacks; (16) Heliocentric
  velocity from the literature; (17) list of references:
  K99: \citet{kp99}; M04: \citet{mieske04}; F07: \citet{firth07}; B07:
  \citet{bergond07}; M08: \citet{mieske08}; G09: \citet{gregg09}; S10:
  \citet{schuberth10}; P18: \citet{pota18}.  The full table is
  available in electronic form at the CDS.}
\end{sidewaystable}

\subsubsection{GCs and UCDs Selection by shape \& photometric properties}

At the assumed distance of Fornax, our best resolution of
$FWHM_{a}\sim0\farcs 7$ (e.g. field FDS\#1 $a$-stack) corresponds to a
physical size of $\sim68$ pc.  Using specific analysis tools
\citep[e.g. Baolab,][]{larsen99}, sources down to $\sim FWHM/10$,
$\sim7pc$ for us, are marginally resolved, and can be analyzed and
identified as slightly resolved sources.  Typical GC half light radii
of 2-4 pc are found in Fornax GCs from high-resolution ACS data
\citep[][]{jordan09,masters10,puzia14}. Using as reference the catalog
of Fornax GC candidates by \citet{jordan15}, $\sim0.5 \% $ of the best
sample ($p_{GC}\geq0.75$) has an half light radius $r_h\geq7$ pc
estimated in both $g$ and $z$ bands. Hence, even at the best
resolution, we can assume the largest fraction of GCs in our catalogs
are indistinguishable from point like sources.

To identify compact stellar systems we adopted a procedure similar to
our previous works \citep{cantiello18vegas,cantiello18n253}.  We
relied on several indicators of compactness derived from the
multi-band $a$-stacks, as on such frames we have the lowest
field-to-field variation, and -by construction- the best seeing over
the entire FDS and FDSex areas. As in the previous works, we combined
the selection based on $CI_n$ to other morphometric indicators from
DAOphot and SExtractor (elongation, flux radius, FWHM, class star,
sharpness). This refines and further cleans the final sample of
compact sources by the possible outliers not identified by using the
sole $CI_n$, or by any other single indicator.

   \begin{figure*}[ht]
   \centering
   \includegraphics[width=17cm]{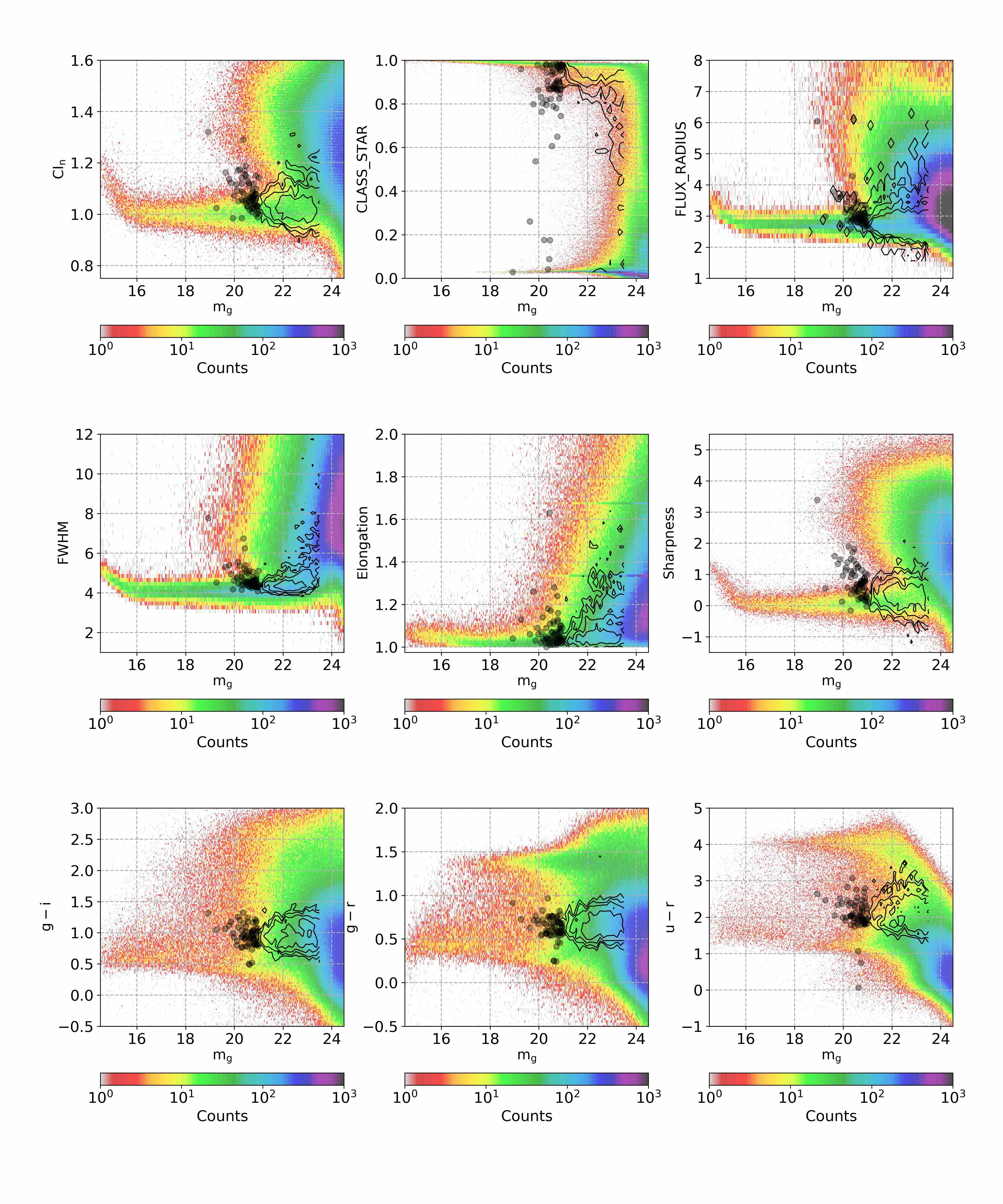}
      \caption{Hess diagrams of several morphometric and photometric
        indicators used to select GC candidates, with overlaid the
        contour levels of GCs in the master reference catalog, and
        UCDs (black circles).}
         \label{selections}
   \end{figure*}

A comparison of the $CI_n$ distribution for the full $ugri$ sample and
for the GCs in the master catalog is shown in Figure \ref{selections}
(upper left panel).  From the comparison with the reference sample
(dark contour levels in the panel) we find that the GC locus extends
over the $CI_n\sim1$ line, with a tail toward larger $CI_n$ values at
fainter $m_g$ magnitudes. UCDs are also reported in the figure, with
black filled dots, and show small but noticeable offsets with respect
to the median properties of confirmed GCs, in particular for the
  size-dependent parameters (like flux radius and FWHM). Such an
effect depends on the evidence that UCDs can have effective radii a
factor of several times larger than GCs
  \citep{mieske08,misgeld11}, i.e. they appear resolved, or slightly
resolved, in our multi-band best seeing image stacks.

In Figure \ref{selections}, we also show some of the other
indicators used to identify GCs, together with UCDs and contour levels
of the master catalog for the appropriate diagram.

To define the best GC selection intervals for each indicator, we
analyzed the master catalog using GCs brighter than $m_g=23.5$, and
derived the median and the $rms_{MAD}$ for each indicator. The results
are reported in Table \ref{tab_medmad}. In the table we also show the
median properties for the reference sample of 68 UCDs.

%
\begin{table}
\caption{Median properties of the GCs and UCDs in the reference catalog}    
\label{tab_medmad}      
\centering                          
\begin{tabular}{l|rr|rr}
\hline
              & \multicolumn{2}{|c|}{GCs}&  \multicolumn{2}{c}{UCDs}  \\
 Indicator    & Median & $rms_ {MAD}$  & Median & $rms_{MAD}$    \\
\hline
           \gi &     0.93 &     0.13 &   0.94 &   0.12 \\
           \gr &     0.63 &     0.09 &   0.64 &   0.08 \\
           \ur &     2.02 &     0.26 &   2.03 &   0.21 \\
       $CI_n$ &      1.03 &     0.03 &   1.06 &   0.03 \\
 $CLASS\_STAR$ &    0.96 &     0.03 &   0.88 &   0.08 \\
         FWHM ($\arcsec$)&     0.94 &     0.08 &   0.91 &   0.04 \\
 Flux Radius ($\arcsec$) &     0.55 &     0.03 &   0.60 &   0.03 \\
   Elongation &     1.09 &     0.05 &   1.04 &   0.02 \\
    Sharpness &     0.32 &     0.26 &   0.69 &   0.34 \\
     $N_{sources}$ &  \multicolumn{2}{|c|}{2138} & \multicolumn{2}{c}{68} \\
  
\hline
\end{tabular}
\end{table}

In addition to morphology, we refine the catalog of candidate compact
sources by their photometric characteristics: the shape of the GCLF
(or the magnitude interval for known UCDs), the color intervals, and
the errors on colors.

In our previous works, mostly focused on NGC\,1399, we adopted as
bright magnitude cut to the GCLF the magnitude $3\sigma_{GCLF}$ above
the turn-over $m_{g}^{TOM}$ of this bright cD galaxy at the
photo-center of Fornax. The Fornax cluster, with an estimated total
line of sight depth of $\sim2$ Mpc \citep{blake09}, has member
galaxies located at different physical distances.  Adopting the ACSFCS
results, the median $g$-band GCLF turn-over magnitude and
$\sigma_{GCLF}$ values are $m_{g}^{TOM}=24.03\pm0.15$ mag and
$\sigma_{GCLF}=0.94\pm0.11$ mag \citep{jordan07}.  A $3\sigma_{GCLF}$
cut above the median TOM corresponds to $m_g\sim21.2$ mag.  For a
rough estimate of the number of GCs lost with such bright cut level,
we again take as reference the ACSFCS full list of GCs hosted by 43
Fornax galaxies \citep{jordan15}. The list contains 53 GCs brighter
than $m_g=21.2$ mag ($\sim0.8\%$ of the sample\footnote{Some even
  brighter GCs are missed in the ACSFCS, as shown by
  \citet{fahrion19}.}). Hence, in what follows we assume $m_g=21$ mag
as bright cut of the GCLF, which includes 99.5\% of the likely GCs
sample in the ACSFCS sample.  The bright cut helps having a sample of
GC candidates with lower stellar contamination, at the cost of an
expected minimal impact on the GC population.  We will in any case
also analyze candidates within $19.0 \leq m_g \leq21.0$ mag, the
magnitude interval corresponding to UCDs in Fornax
\citep{mieske12}. These systems share many characteristics with GCs
but, as mentioned above, have larger effective radii than GCs (see
Figure \ref{selections}).

As maximum color uncertainty we choose $\Delta(g{-}i)_{max}=0.15$ and
$\Delta(u{-}r)_{max}=0.3$, corresponding approximately to half of the
separation between the blue and red peaks of the GCs color
sub-populations host in typical bright galaxies
\citep{cantiello18vegas}.

Thanks to the multiple color coverage, the selection of candidates can
be improved using color-color criteria, rather than {\it flat}
single-color ranges. The contour levels in the color-color diagrams of
Figure \ref{hess_histo} reveal relatively narrow color-color loci of
GCs.  A simple color-color selection box (e.g. black dotted lines in
the upper left panel of the figure) would imply a trivial
contamination from either stars or background objects. Instead, we
proceed by inspecting in the color-color planes all sources satisfying
the morpho-photometric parameters identified above. Finally, only the
sources inside the color-color contours of the reference sample are
identified as candidates and used for further analysis (see next
section).

\noindent
In summary, to identify the least contaminated and most complete
possible GCs (and UCDs) catalog from our photometry, we adopted a
three steps strategy.  First, we generated a master GCs (and UCDs)
catalog using confirmed sources in the literature. From the GCs
catalog we cut out all sources fainter than $m_g=23.5$ mag, to better
identify the morpho-photometric loci of GCs; the cut is adopted only
for the reference catalog, for the GC identification and analysis on
the FDS catalogs we will adopt a $\sim1$ mag fainter limiting
magnitude to increase the sample of GC candidates (see below). Second,
we used the control parameters shown in Fig. \ref{selections}, and the
properties of the master catalogs to define the best intervals for GCs
and UCDs selection. These selection criteria are then independently
applied to the FDS and FDSex catalogs. For some parameters we adopted
as confidence intervals the ranges from the master catalogs, using the
$median\pm N\times rms_{MAD}$, with $N=4/2$ for GCs/UCDs respectively
(median and $rms$ from Table \ref{tab_medmad}); for the GCLF, colors,
and color errors, we proceeded as described above. The complete list
of parameters, together with the used ranges, is reported in Table
\ref{tab_params}. Third, the sample of compact sources after the
previous steps was inspected in the color-color plane to further
narrow down the contamination using the contour levels derived from
the master catalog.


\begin{table}
\caption{Photometric and morphometric parameters adopted for source selections}    
\label{tab_params}      
\centering                          
\begin{tabular}{l|rr|rr}
\hline
              & \multicolumn{2}{|c|}{GCs}&  \multicolumn{2}{c}{UCDs}  \\
 Indicator    & Min. & Max.     & Min. & Max.    \\
\hline         
$m_g$           &      21.0          &       24.5     &      19.0    &         21.0  \\
\gi             &      0.5         &       1.4        &      0.5        &       1.4     \\
           \gr  &      0.25        &       1.1        &      0.25       &       1.1     \\
           \ur  &      1.2         &       3.4        &      1.2        &       3.4     \\
       $CI\_n$  &      0.90        &       1.17       &      1.00       &      1.13     \\
 $CLASS\_STAR$  &      0.50        &       1.00       &      0.5        &       1.00    \\
         FWHM   &      0.61        &       1.26       &      0.8        &       1.12     \\
  Flux Radius   &      0.42        &       0.68        &     0.5        &       0.8     \\
   Elongation   &    $\nodata$     &       1.30       &    $\nodata$    &       1.5     \\
   Sharpness    &     -0.75        &       1.40       &      0.3        &       2.0     \\
$\Delta \gi$    &    $\nodata$     &       0.15       &    $\nodata$    &       0.15    \\
$\Delta \ur$    &    $\nodata$     &       0.30       &    $\nodata$    &       0.30    \\
\hline
\end{tabular}
\end{table}

\subsection{Surface distribution of compact sources over the FDS area}

The analysis of the GCs over the FDS and FDSex area, together with the
comparison with similar datasets, will be presented in more
  detail in a forthcoming paper. In what follows we show a
preliminary determination of GCs and UCDs surface density maps as an
example use of the FDS catalogs, based on the source selection
strategies described in the previous section; in Section
\S\ref{ctortora} we will also show an example of use of the catalogs
for the study of background galaxies.

\subsubsection{Globular clusters and UCDs distribution maps}

Using the identification scheme described above, we inspect the GC
distribution maps over the FDS and FDSex areas using as reference the
$ugri$ and $gri$ selections, respectively.

GC candidates are derived by cross-matching the color-color regions of
pre-selected GC candidates (Table \ref{tab_params}), with the
color-color loci of GCs identified in the master sample. Candidates
falling in the contour levels of higher GCs density in the two-color
diagram have higher likelihood of being true GCs. However, the narrow
color-color range also implies lower completeness. In what follows,
then, we analyze the GC density maps for candidates over different
color-color contour levels.

\begin{figure*}
   \centering
   \includegraphics[trim={4.5cm -0.5cm 5cm 2cm},clip,scale=0.073]{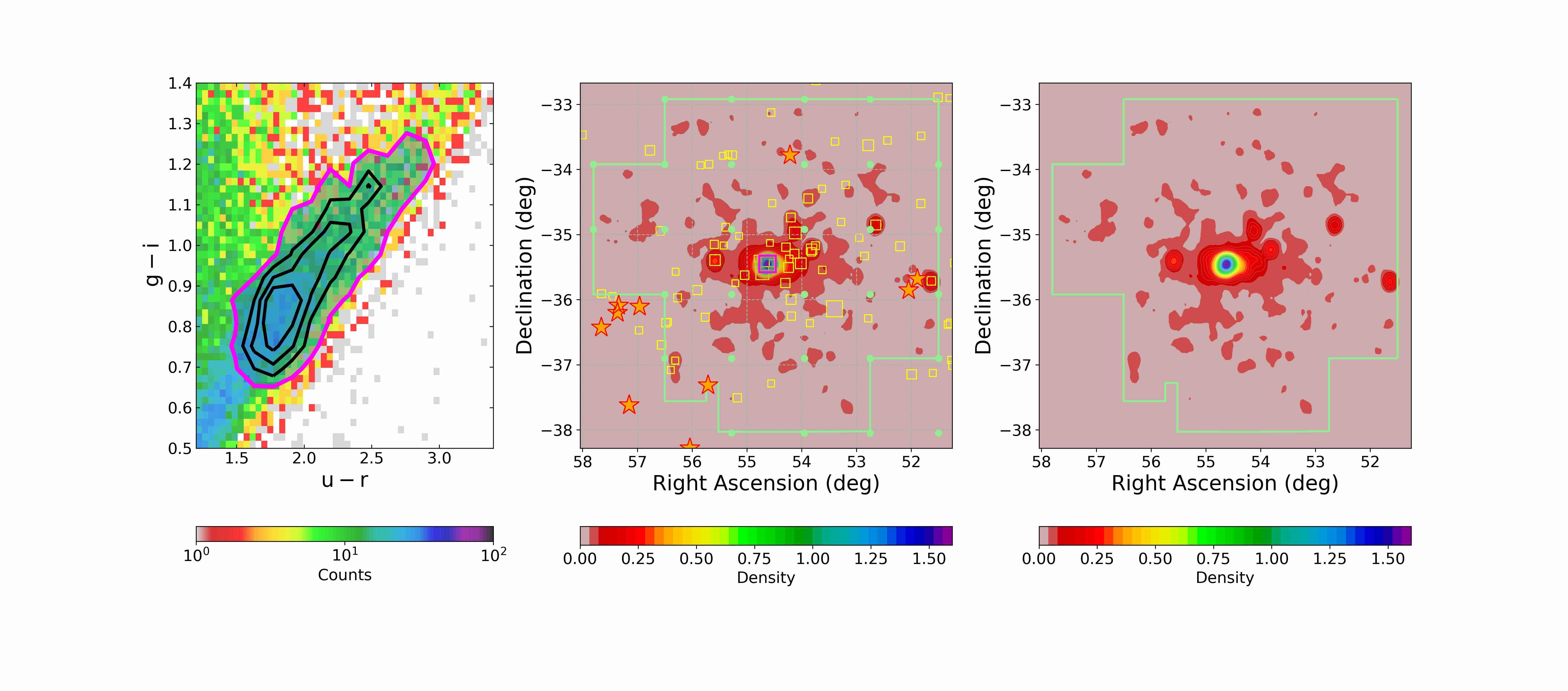}
   \vskip -1.25cm
   \includegraphics[trim={4.5cm -0.5cm 5cm 2cm},clip,scale=0.073]{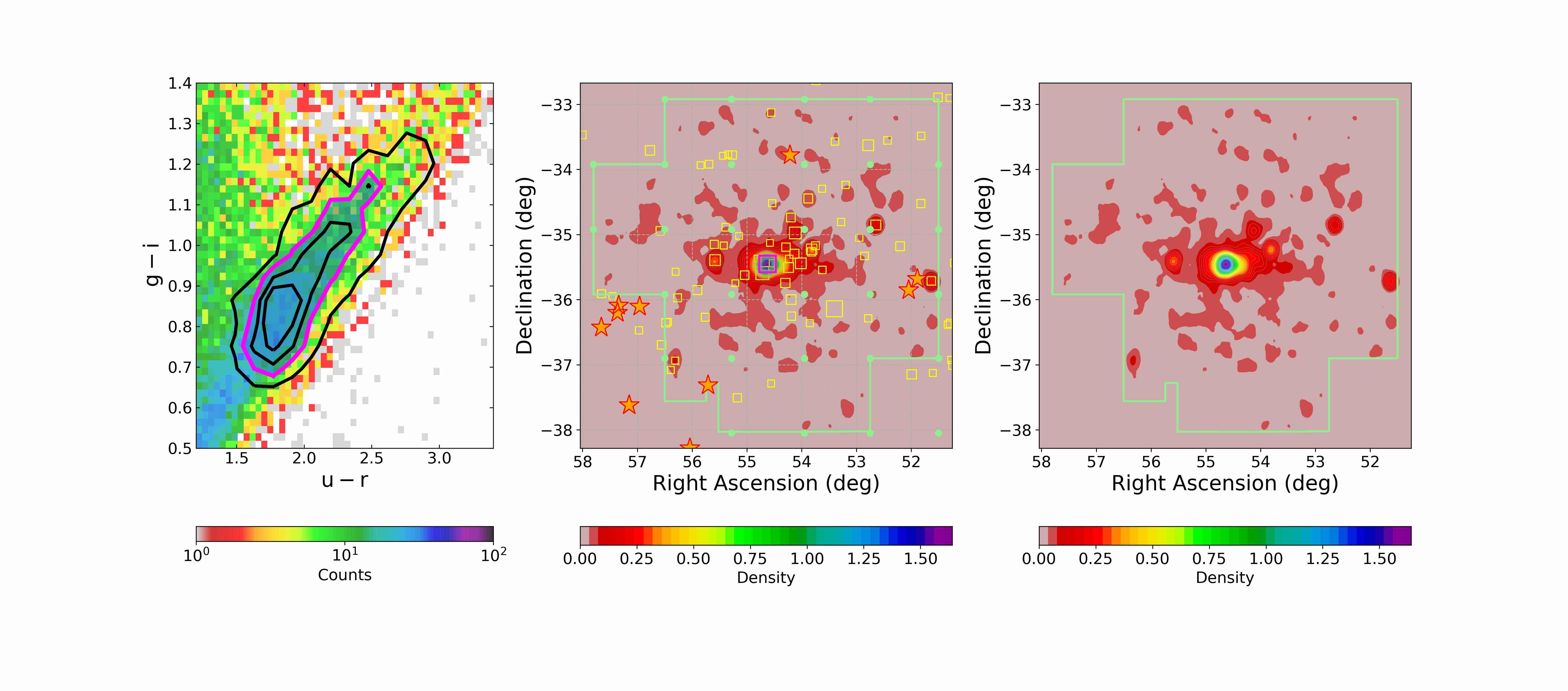}
   \vskip -1.25cm
   \includegraphics[trim={4.5cm -0.5cm 5cm 2cm},clip,scale=0.073]{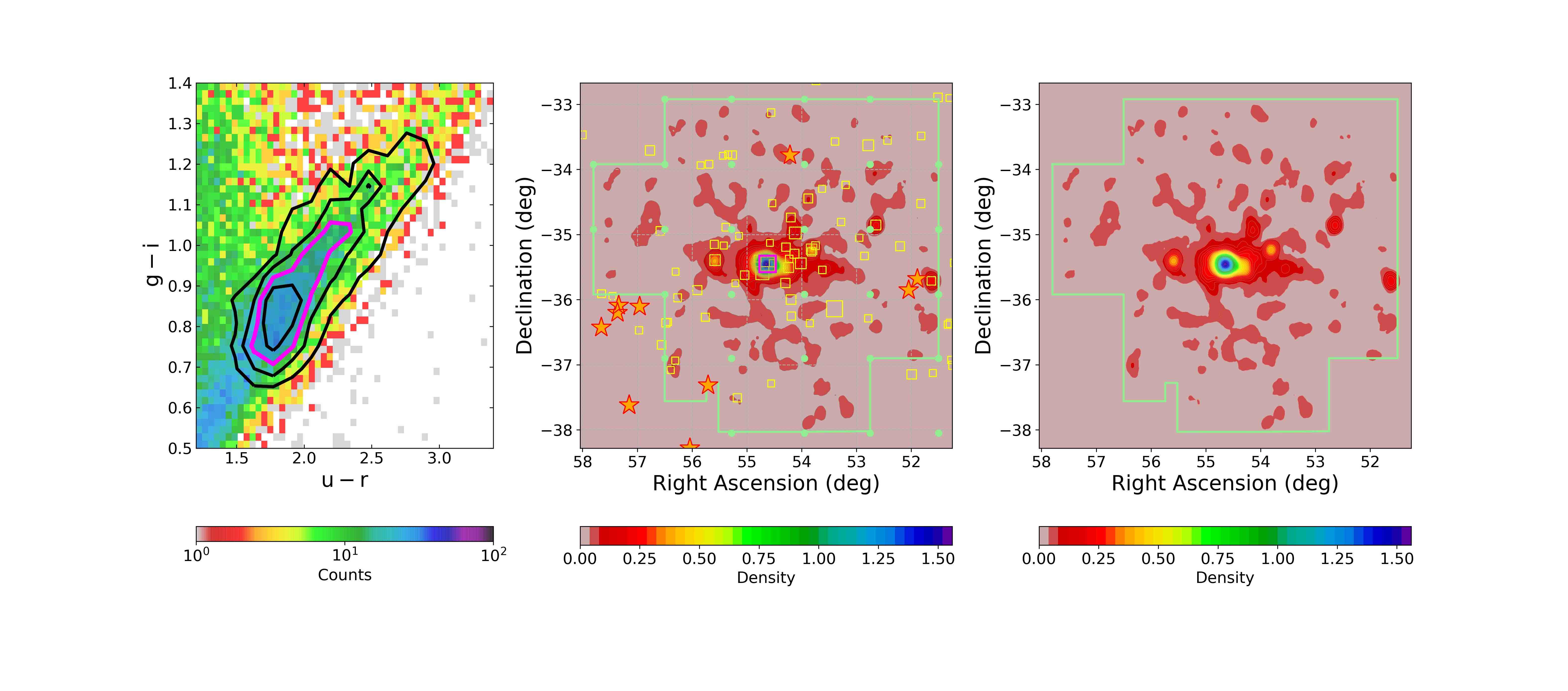}
   \vskip -1.25cm
   \includegraphics[trim={4.5cm -0.5cm 5cm 2cm},clip,scale=0.073]{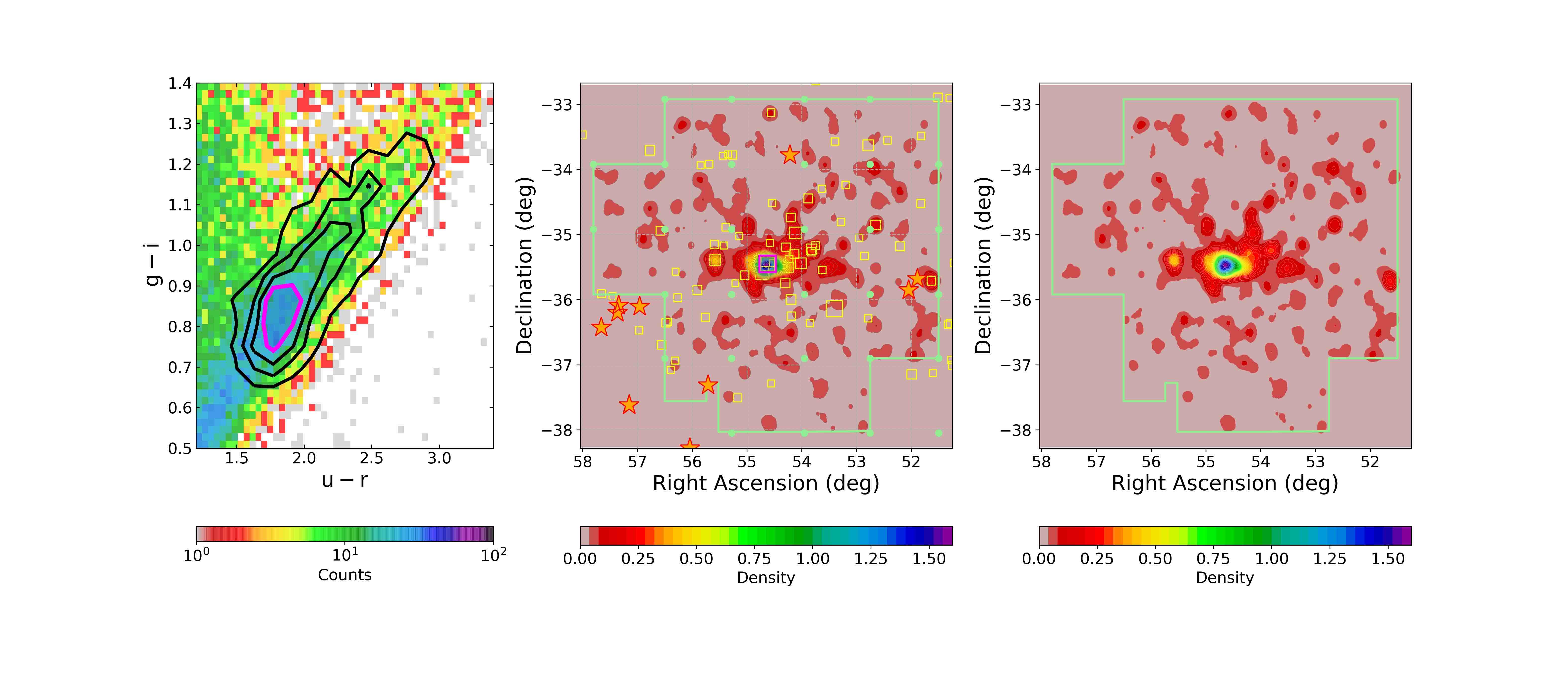}
   \vskip -1.1cm
      \caption{\tiny{Surface density maps of GC candidates over the
          FDS area. Upper left panel: Color-Color Hess diagram for GC
          candidates selected using the parameters in Table
          \ref{tab_params}. The contour lines refer to the master GC
          sample. All GC candidates in the color-color contour level
          shown with a thick magenta solid line (also evidenced with a
          gray shaded area) are used for the density maps in the
          middle and right panels.  Upper middle panel: Density map of
          the GC candidates within the shaded area highlighted in the
          left panel. The density is in number of candidates per
          square arcmin. East is left, north is up. The light green
          line shows the FDS footprint; filled green dots mark the
          limits of single pointings; five pointed stars mark stars
          with $m_V\leq7$ mag; yellow squares show galaxies brighter
          than $B_T=16$ mag, with symbol size scaled to galaxy total
          magnitude; NGC\,1399 is also marked with a magenta empty
          square. Upper right panel: as upper middle panel, except
          that all reference sources and lines are not plotted to
          highlight the GC structures in the area. Second to fourth
          row of panels: as upper row, but for the other narrower
          contour levels of the color-color diagram, as evidenced with
          the magenta contour in the first column of panels.  From
          upper to lower panels the number of GC candidates within the
          color-color region identified with red contour level is:
          5.650, 3.650, 2.170, and 900, respectively.}}
         \label{mappa_gc}
   \end{figure*}

Figure \ref{mappa_gc} shows the two-dimensional projected distribution
over the $\sim21$ sq. degree area of FDS.  In the left panels of the
figure we plot the color-color Hess diagrams of all sources identified
with the selection criteria in Table \ref{tab_params},
  overplotting the contour levels of the GCs in the master
sample. Even after all morpho-photometric cleaning of the sample
(except for the color-color selection), a substantial fraction of
selected candidates lies outside the expected GCs color-color region
identified by the contour levels in the panel.

The middle and right panels of Figure \ref{mappa_gc} show the maps of
GCs identified adding also the color-color contour level selection,
i.e. of all sources falling in the contour levels marked in the left
panels of Figure \ref{mappa_gc}. Each row of panels in the figure
refers to a different contour level, indicated by the thick magenta
contour in the left panel. Again, the inner contours pinpoint regions
with higher GCs density in the color-color diagram, thus the level of
contamination from non-GCs decreases in the maps from the upper to
lower panels in Figure \ref{mappa_gc}; vice-versa, because of the
smaller color-color intervals, lower panels suffer for higher
incompleteness fractions.  In particular, the lowermost panel is
limited to a blue color-color region, hence mostly representative on
the blue-GCs sub-population, also discussed below.

We calculate the smooth density maps using non-parametric kernel
density estimates based on FFT convolution\footnote{We used the KDEpy
  python 3.5+ package, which implements several kernel density
  estimators. See the web pages of the package for relevant
  literature:
  \url{https://kdepy.readthedocs.io/en/latest/API.html\#fftkde}}.
After various tests, we adopted a grid mesh size of $\sim0.1\arcmin$
spacing, smoothed with an Epanechnikov kernel, with kernel
bandwidth\footnote{Using a Gaussian kernel, the bandwidth is
  equivalent to the $\sigma$ of the distribution.} five times the grid
size.

Although obvious differences appear between GC maps drawn from the
diverse color-color contour levels, there are several recurrent
patterns appearing at various levels of selection, that is at
different levels of GC contamination and incompleteness. The
recurrence of the sub-structures over various GC color-color
  contours supports the reality of the sub-structure itself. Some of
these patterns were also discussed in our works \citep{dabrusco16,
  cantiello18vegas}, over a smaller survey area and using
  partially different data and algorithms; yet here we observe
several new features, that are possible extensions to the ones
described previously.

\begin{figure*}
   \centering
   \includegraphics[trim={-2cm 3cm 1cm 1cm},clip,scale=0.75]{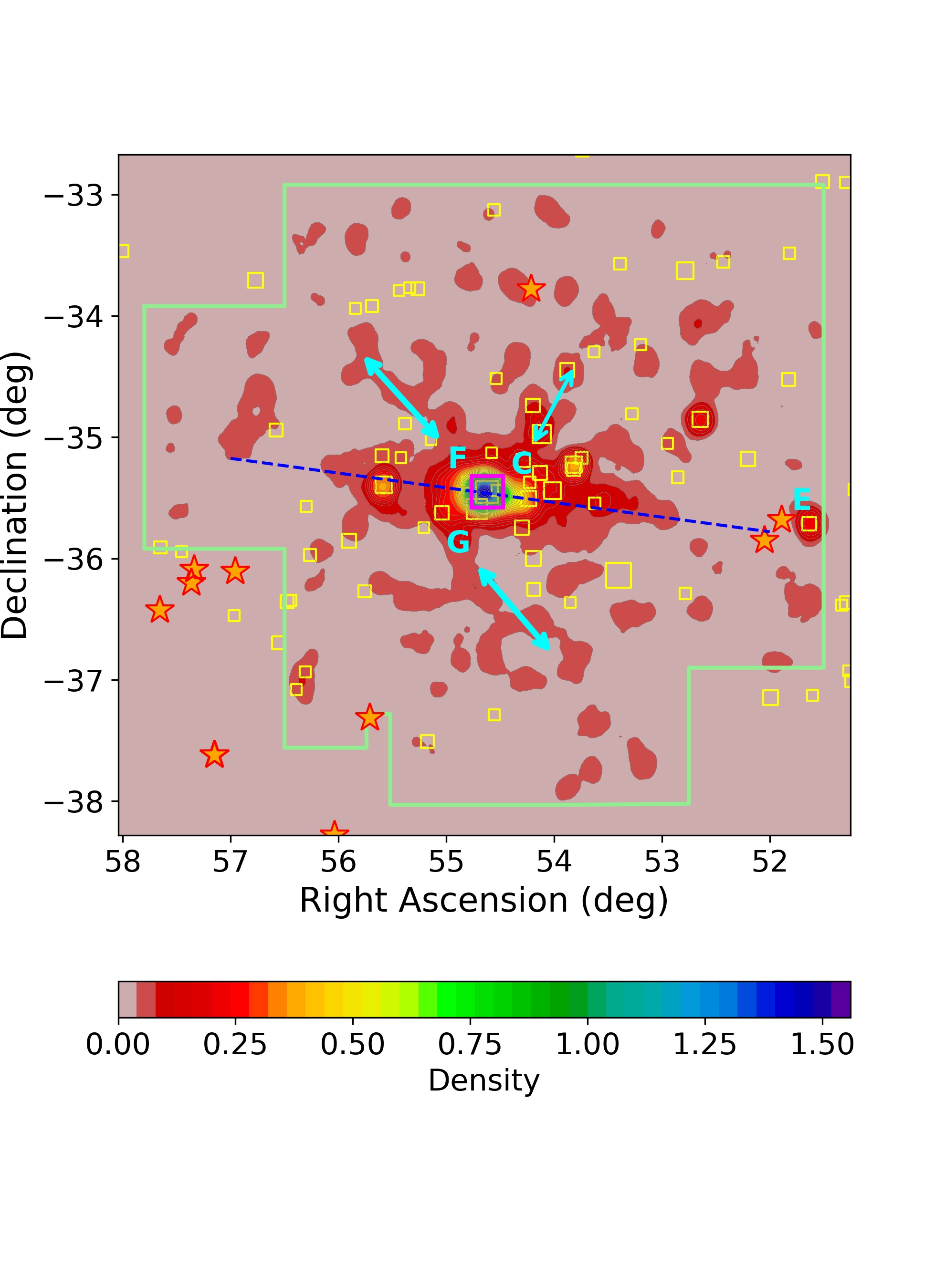}
       \caption{Single panel view of the 2-D GC surface
         distribution. Iso-density contours and symbols are the
         same as in Figure \ref{mappa_gc} (third contour level
         plots).  Light blue arrows and labels indicate the GC
           overdensities discussed in the text. The blue dashed line
           shows the $\sim10^{\circ}$ tilt in the direction of
           NGC\,1336 ('E' label in the figure).} 
         \label{mappa_gc_single}
   \end{figure*}

{\it $\bullet$ Central over-density:} For sake of clarity, in Figure
\ref{mappa_gc_single}, we plot the density map relative to the
  third contour plot (third row in Figure \ref{mappa_gc}).  The
peanut shaped distribution of GCs, elongated in the E-W direction of
the cluster, with a marked peak on NGC\,1399, was already found in our
studies relying on data of the central FDS area, within $52.5 \leq
R.A.~(deg) \leq 56.5$ and $-37 \leq Dec.~(deg)\leq-35$ (a total of
$\sim7.5$ sq. degrees).

In the new dataset, covering $\sim4$ times the area previously
inspected, we find a $\sim10$deg tilt of the position angle for the
broad distribution of inter-galactic GC candidates, tilting in the
direction of \object{NGC\,1336} (the tilt direction is also indicated with
  a blue dashed line in Figure \ref{mappa_gc_single}). The length of
the last isodensity contour is $a=2.6\pm0.2~deg$ (or $920\pm60$ kpc),
obtained combining the sizes from the four maps in Figure
\ref{mappa_gc}.  The width of the distribution is of
$b=0.89\pm0.03~deg$ (or $310\pm10$ kpc), implying an ellipticity
$\epsilon=1-b/a\sim0.65$, slightly larger than what previously found
on smaller scales \citep{kim1399, cantiello18vegas}.

{\it $\bullet$ $F$ \& $G$ features:} In the distribution, besides the
obvious case of NGC\,1399 and its fainter close companions, we observe
several regions of marked over-density in correspondence with bright
galaxies or pair of galaxies: \object{NGC\,1427}, \object{NGC\,1374/1375}, \object{NGC\,1351},
all with $B_T\leq12$ mag, and of NGC\,1336, which is $\sim1.5$ mag
fainter than the others.  The GCs peaks on these regions were already
commented in our previous works. However, thanks to the larger area
analyzed and the different detection strategy the new photometric
sample reaches $\sim1$ mag deeper $u$-band, we now find that such
structures are connected and extend to larger clustercentric
radii. The $F$ and $G$ features described in \citet{dabrusco16}
(arrows in Figure \ref{mappa_gc_single}) extend $\sim1.5$ degrees
($\sim 0.5$ Mpc) South-West and North-East of the cluster core,
respectively.  These substructures do not cross any galaxy brighter
than $B_T=16$ mag, both overlap a handful of galaxies with $16\leq B_T
\leq18$ mag (absolute magnitude $-15.5\leq M_{B,tot}~(mag)\leq-13.5$),
and a dozen of fainter galaxies, down to $B_T\sim20$ mag ($
M_{B,tot}=-11.5$ mag). The $F$ extension, points toward a group of
five galaxies with magnitudes $B_T$ between 13.5 and 16 mag, dominated
by ESO\,358-050, where no GC structure or overdensity is noticeable in
any of the GC color-contour selections.

The level of persistence of the $F$ and $G$ structures changes with
the selection contours.
To estimate the level of significance of both these overdensities we
proceed as follows. First, taking as reference the third contour level
in Figure \ref{mappa_gc}, we count the number of GC candidates in the
$F$ and $G$ feature density contours ($N_{X}$, with $X$ referred to
the $F$ or $G$ region). Then, to define a background level, we move
the same density contours around the FDS area, avoiding the central
overdensity and the regions with galaxies brighter than $B_T\sim16$,
and count the number of candidates in such regions. For each feature,
we identified seven independent regions for background estimation over
the survey area; then we used the median and $rms_{MAD}$ of the GC
number counts in the seven regions ($N_{X,back.}$, $rms_{X,back.}$) to
quantify the $F$ and $G$ overdensity ratio as follows:
\\ $\Sigma[(in{-}out)/err]=(N_{X}-N_{X,back.})/(rms_{X,back.}^2+\delta
N_{X}^2)^{1/2}$. \\

By definition, $\Sigma[(in{-}out)/err]$ quantifies the ratio between
the difference of counts in and out the $X$ feature, and the squared
sum of the standard deviation of both counts, assuming a poissonian
fluctuation for $N_{X}$ ($\delta N_{X}=N_{X}^{1/2}$). We obtain
$\Sigma[(in{-}out)/err]\sim4.2$, for $F$ and $\sim4.4$ for $G$,
meaning that the GC candidates overdensity with respect to the diffuse
background GCs component, is at least factor of four larger than the
estimated total expected counts fluctuation in both regions.  A
similar result, although on smaller regions, with a different
(shallower) sources catalog and with independent algorithms, was found
by \citet{dabrusco16}.

The $F$ is more evident in the wider color-contours selections (upper
two panels in Figure \ref{mappa_gc}), which also include the red GCs,
that are mostly expected to be closely bound to the galaxies; because
of the wider selection intervals, this feature is also likely to have
higher fore/back-ground contamination. The $G$ structure, instead,
appears more connected to the blue GC population (lowermost panel in
the figure); the properties of such coherent structure extending over
cluster scale, over an area devoid of bright galaxies and composed
mostly of blue GCs --the GC sub-population typically found in the
outer galactic regions-- suggest its inter-galactic nature. We
speculate that the $G$ feature might be connected with \object{NGC\,1404}, as a
stream of blue GCs possibly leading/tailing from the galaxy; the galaxy has an
overall $z$-band specific frequency $S_{N,z}=0.30\pm0.00$, and within
one effective radius $S_{N,z,In}=0.12\pm0.01$ \citep{liu19}. The whole
median of the ACSFCS sample is $\langle S_{N,z}\rangle=0.82\pm0.37$,
or $0.93\pm0.26$ if limited to the five brightest galaxies in the main
Fornax cluster after excluding NGC\,1399 and NGC\,1404
itself\footnote{The median with NGC\,1399 and NGC\,1404, doesn't
  change notably, being $0.93\pm0.41$.}; for the $S_{N,z,In}$ from the
combined Fornax and Virgo cluster sample \citep[Table 4 in][]{liu19},
and limited to galaxies brighter than $M_z\sim-20.7$ mag, we obtain
$\langle S_{N,z,In}\rangle=0.32\pm0.18$. Hence, in all cases NGC\,1404
is a noteworthy case of bright galaxy with a GC population
consistently lower than average. \citet{bekki03} have a dynamical
model for the GCs system of NGC\,1404, explaining its low specific
frequency as an effect of the tidal stripping of GCs by the
gravitational field around cluster core, dominated by NGC\,1399.  The
authors find that, at given models input conditions (highly eccentric
orbit, initial scale-length of the GCs system twice as large as the
galaxy effective radius) NGC\,1404 GCs population can be reduced
through stripping to the presently observed value.  One of the
observable characteristics predicted by \citeauthor{bekki03} is the
formation of an elongated or flattened tidal stream of GCs.

Furthermore, the complex structure of the Fornax X-ray halo
\citep{paolillo02,su17} has been explained by \citet{sheardown18}
using hydrodynamics simulations, by the orbital motion of NGC\,1404
within the cluster, assuming that the galaxy is at its second or third
passage through the cluster center.

{\it $\bullet$ NGC\,1336:} The new photometry confirms the peculiarity
of NGC\,1336 with respect to the rest of the cluster: we find its GCs
overdensity \citep[$E$ feature in][]{dabrusco16} isolated with respect
to the rest of the cluster-wide GCs system.  The distinctiveness of
NGC\,1336 is also discussed by \citet{liu19}, who find that it has the
second highest GC specific frequency, after NGC\,1399, and the largest
3D clustercentric distance in the ACSFCS sample. The relative
isolation of the galaxy from the Fornax core, at $\sim 2$ times of the
cluster virial radius, also supported by the lack of GC streams toward
the core, might strengthen the hypothesis by \citeauthor{liu19} that
it {\it ''is an infalling central galaxy with a higher total
  mass-to-light ratio, resembling the behavior of the most massive
  ETGs. Its GC system has possibly experienced fewer external
  disruption processes, and the GCs may have a higher survival
  efficiency.''}  The presence of two kinematically decoupled cores
\citep{fahrion19b}, most probably evidencing a major merger that has
altered the structure of NGC\,1336 significantly, might further
support such hypothesis.

{\it $\bullet$ The $C$ feature:} A further structure, labeled $C$ in
\citeauthor{dabrusco16}, ranges from \object{NGC\,1380} North-West in the
direction of the ringed barred spiral \object{NGC\,1350}. The feature appears
less coherently connected than the $F$ and $G$ in the maps of
Fig. \ref{mappa_gc}, and it crosses four galaxies with $B_T\leq 16$,
thus it might be the result of the projected superposition of several
adjacent GC systems, rather than an intra-cluster GCs structure.

   \begin{figure*}
   \centering \includegraphics[trim={4.5cm 3.5cm 20.5cm 2cm}
     ,clip,scale=0.085]{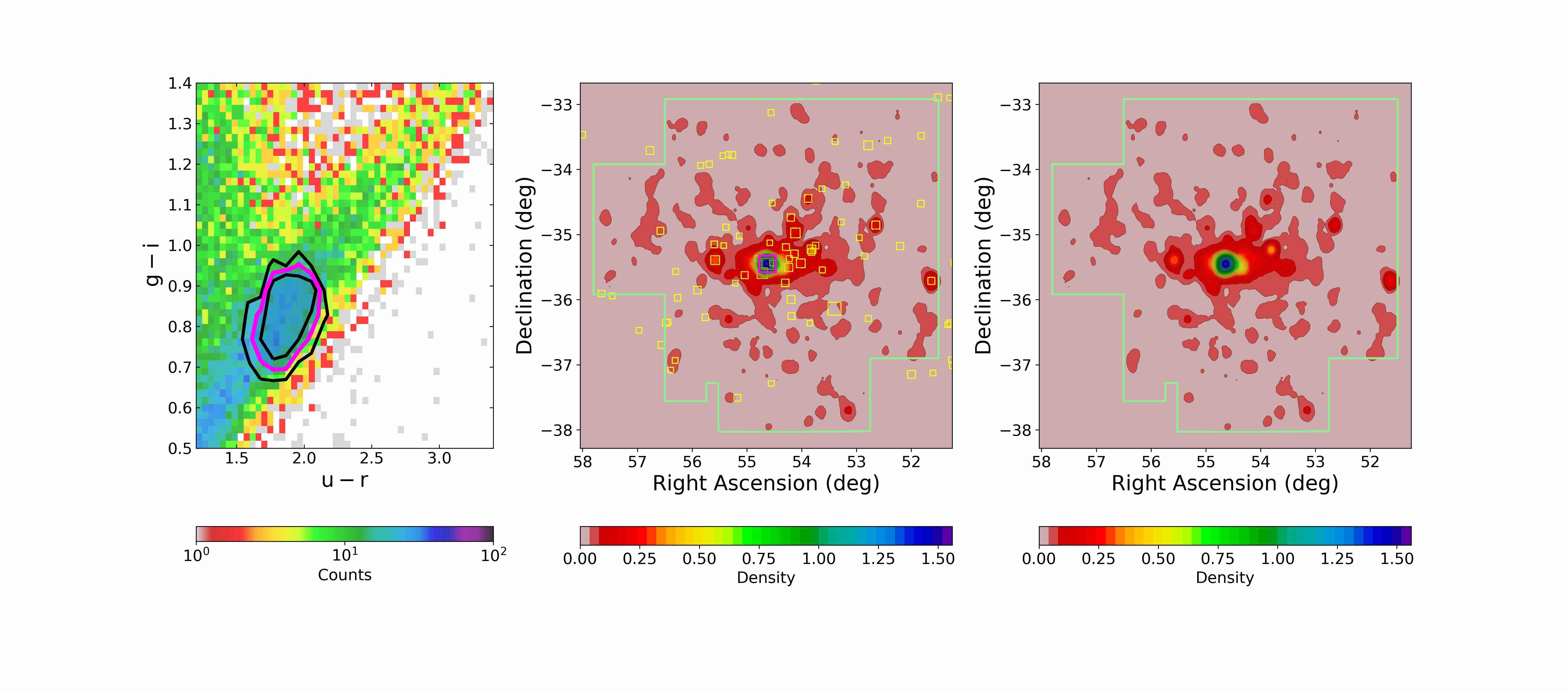}
   \includegraphics[trim={4.5cm 3.5cm 20.5cm 2cm}
     ,clip,scale=0.085]{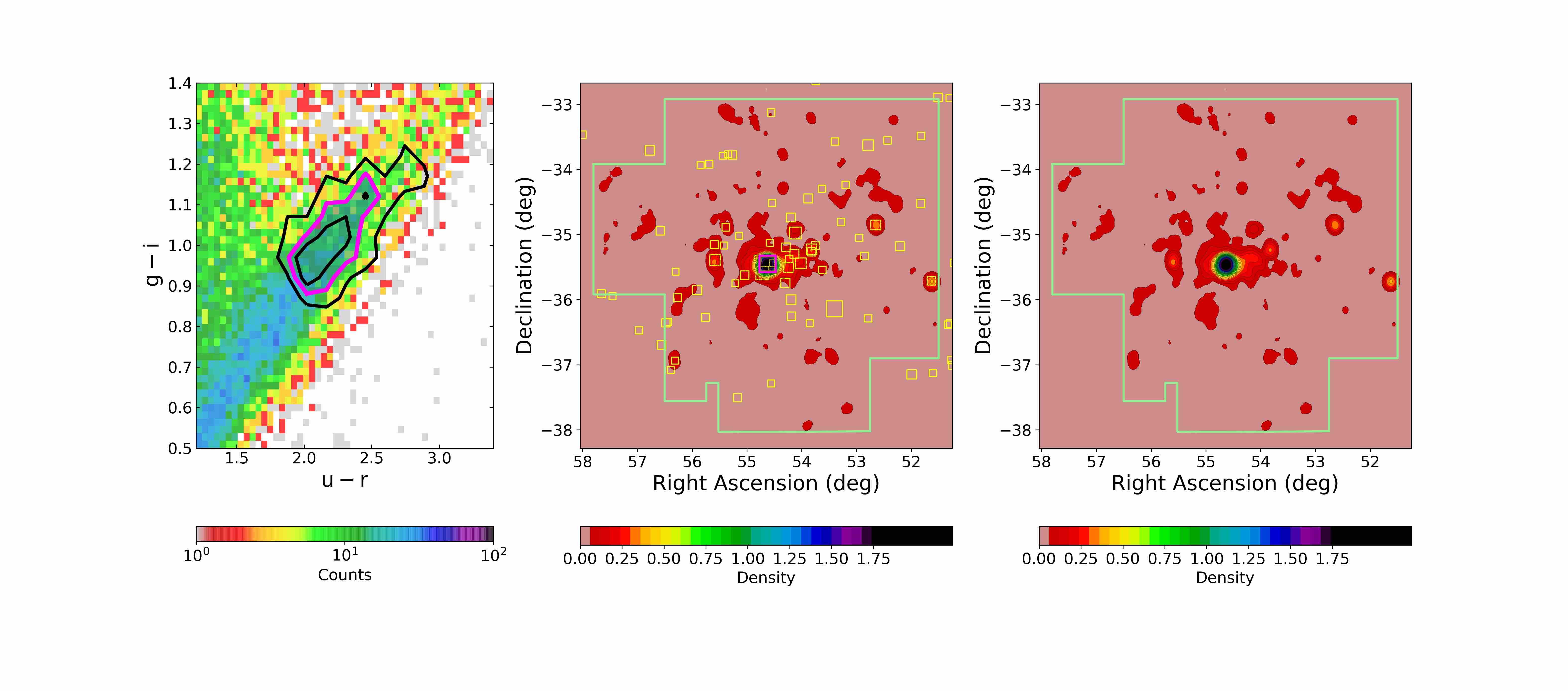}
       \caption{The two-dimensional density maps of blue (upper
         panels) and red (lower panel) GC candidates. Symbols are the same as in Figure
         \ref{mappa_gc}.}
         \label{mappa_brgc}
   \end{figure*}

{\it $\bullet$ Blue and red GCs, foreground stars:} We also plot the
map of blue and red GC candidates in Figure \ref{mappa_brgc}, using
the color-contours shown in the left panels.  To improve the blue/red
GCs separation, taking advantage of the availability of two colors,
the separation between red and blue GC is taken from a linear fit to
the \ur-\gi sequence of the master GC sample, then taking the blue/red
separation from the dip in the distribution projected along this axis,
a procedure we already used in \citet[][see their Figure 7, upper
  panel]{angora19}. The blue/red surface density maps show the
property already anticipated above of red GCs being concentrated on
galaxies, especially on bright ellipticals, and blue GCs covering a
wider area, including the intra-cluster regions.

\begin{figure*}
   \centering
   \includegraphics[trim={4.5cm 3.5cm 5cm 2cm},clip,scale=0.1]{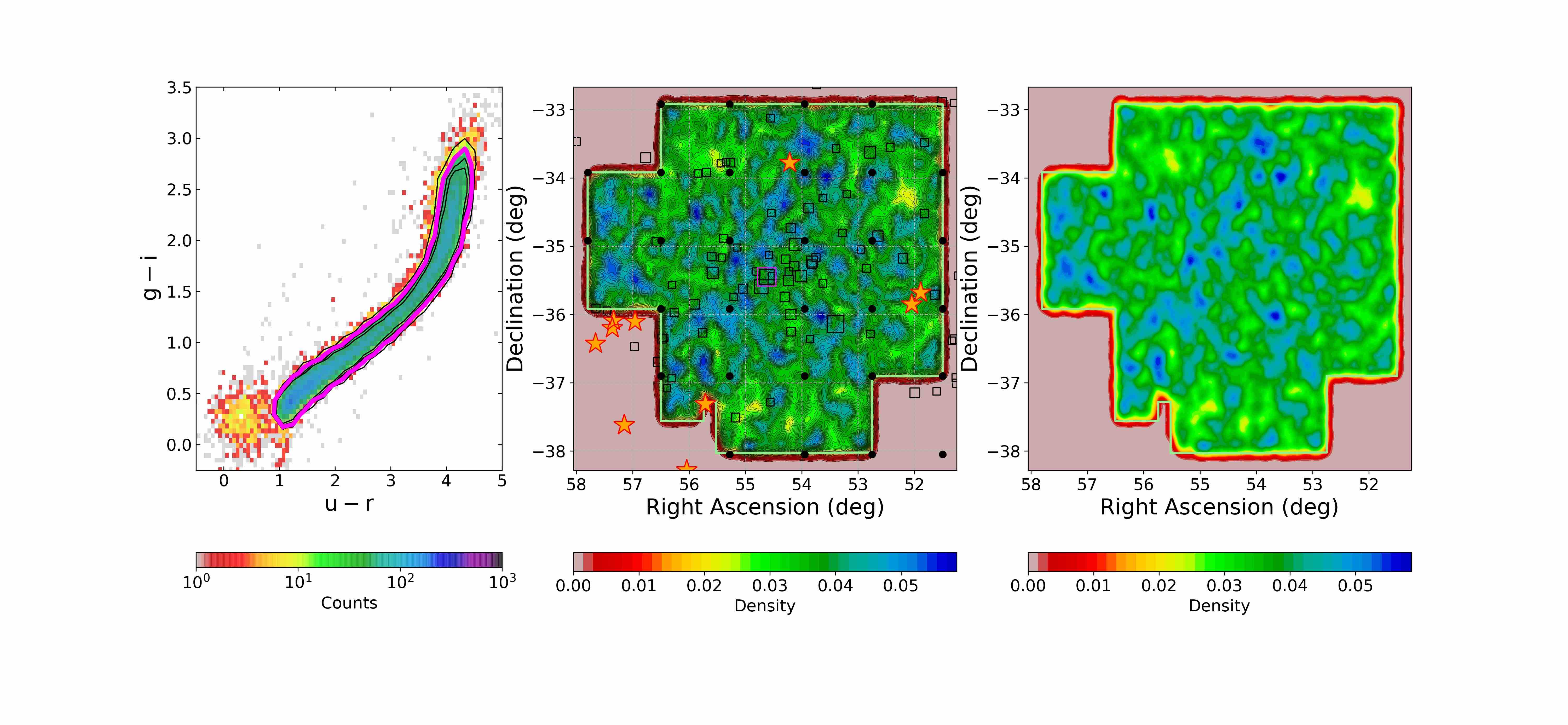}
       \caption{The surface density maps of bright stars. Symbols are the same as in Figure
         \ref{mappa_gc}.}
         \label{star_maps}
   \end{figure*}

For comparison with the previous maps, Figure \ref{star_maps} shows
the stellar density map, where stars are identified as the bright
sources $16\leq m_g~(mag)\leq20.5$, with the same photo-morphometric
properties of GCs (Table \ref{tab_params}) except no color selection
is applied. The stellar map shows both the lack of any obvious
structure over the field, and the large contamination from MW stars:
the map, limited to the brightest part of the field MW stellar
population, is derived from $\sim23.000$ stars, versus the
$\sim5.600/900$ GCs used for the GC maps in Figure \ref{mappa_gc}, and
the $\sim2.200/1200$ blue/red GCs selected for the maps in Figure
\ref{mappa_brgc}.

\begin{figure*}
   \centering
   \includegraphics[trim={4.5cm 3.5cm 5cm 2cm},clip,scale=0.1]{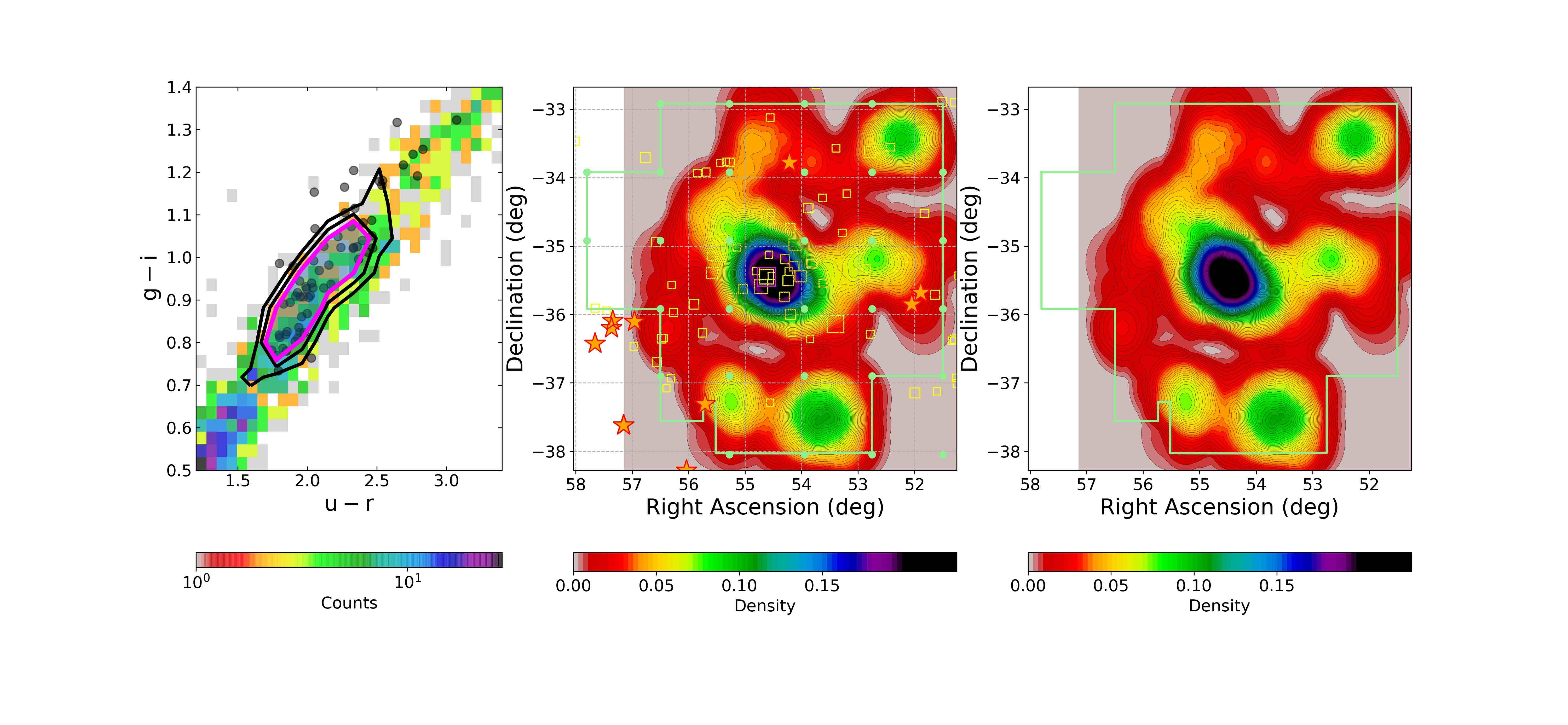}
       \caption{The surface density maps of UCD candidates. Symbols are the same as in Figure
         \ref{mappa_gc}.}
         \label{ucd_maps}
   \end{figure*}

{\it $\bullet$ UCD galaxies:} A further map from the $ugri$ catalog
is shown in Figure \ref{ucd_maps}, with the UCD surface density
distribution over the FDS area, derived using the selection parameters
for UCDs, reported in Table \ref{tab_params}, and the color contours
of known UCDs from the reference sample (magenta solid lines in the
figure).  Unsurprisingly the surface density maps show the
concentration of UCDs rises around the central square degree area of
NGC\,1399. The map is mostly shown for completeness, as number of UCDs
is known to be small, so even a small contamination can significantly
alter the analysis.  With our selection we identify 160 sources, which
probably include a substantial fraction of contaminating stars,
especially in the brightest magnitude bin ($19\leq m_g\leq 20$), and
bright GCs with morphological parameters consistent with the UCDs.
Inspecting separately the maps of bright/faint UCDs candidates,
adopting as separation limit $m_g=20$ mag, we observe that the map for
the faint magnitude bin -- $m_g=20$-$21$ containing 105 candidates at
the given selection criteria -- doesn't change notably with respect to
Fig. \ref{ucd_maps}, and shows an elongated density structure with a
peak close to the cluster core, and two secondary maxima at [RA,
  Dec.]=[53.7, -37.6] and [52.3, -33.5]. The map of the bright
component -- $m_g=19$-$20$, 55 candidates -- does not show any
noteworthy pattern, with sources appearing evenly distributed in the
region, a behavior suggesting large contamination from MW stars in
this magnitude range. The study of the UCD distribution over the area
requires a dedicated analysis to characterize and identify all the
selected UCD candidates, which is beyond the scopes of this study, and
will be addressed in a forthcoming work, also using near-IR photometry
(Saifollahi et al., in prep.).

In conclusion, it is worth highlighting that all the sub-structures
described in this section are relatively insensitive to the main
parameters chosen to identify GC or UCD candidates, and to the details
of the algorithms used to derive the maps themselves, except minor
details which leave unaltered the general presentation above.

\subsubsection{GCs distribution maps over the FDSex area}

\begin{figure*}
   \centering
   \includegraphics[scale=0.45]{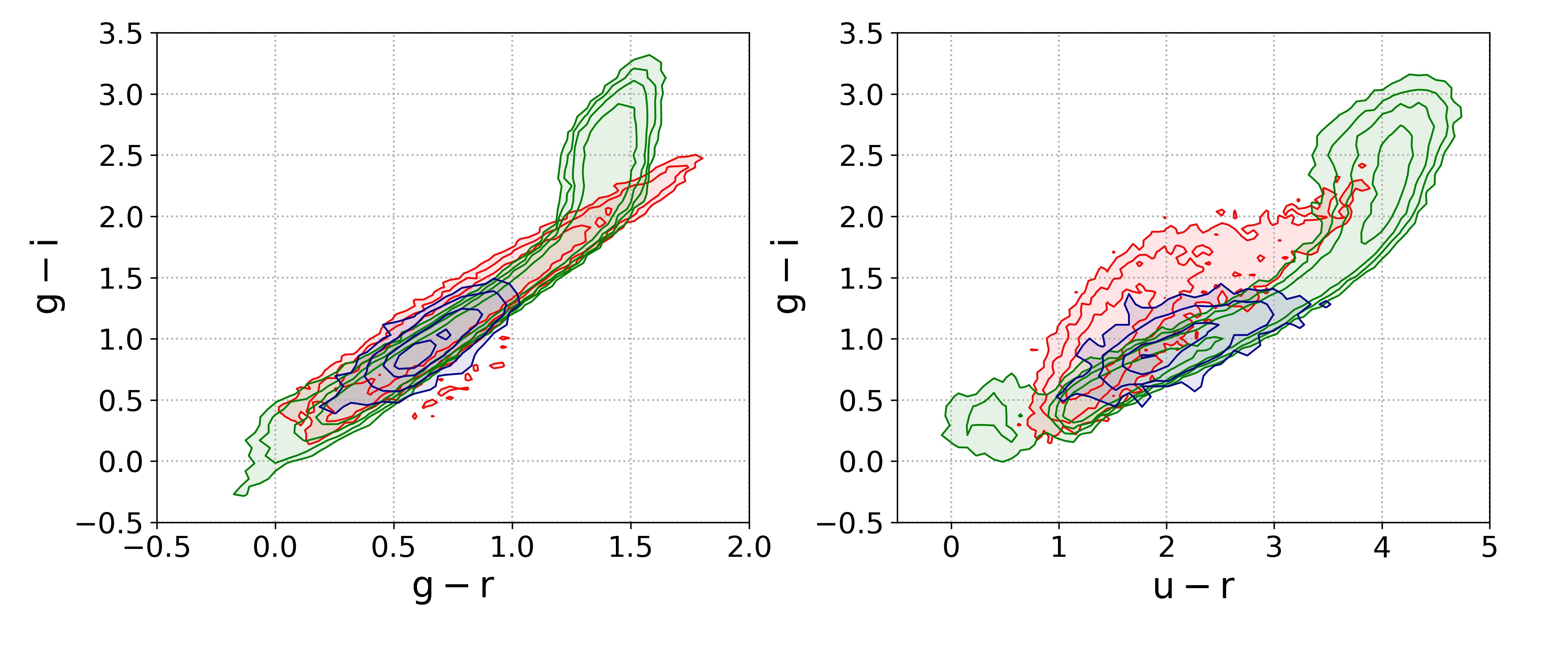}
       \caption{Color-color contour plots of the GC master catalog
         (blue contours and shaded area), of point-like sources
         (green) and of extended sources (red).}
         \label{compactext}
   \end{figure*}

The lack of $u$-band photometry over the FDSex area implies that any
sample of compact sources selected in the area using the same
procedures adopted in the previous section, yet based only on $gri$
photometry, is more contaminated. In Figure \ref{compactext} we plot
the contour levels of the master GC sample (blue lines and shaded
area), and the contour levels of compact (green color, $C.I_n\sim1$)
and extended (red colors, $CI_n\geq1.3$) sources, all brighter than
$m_g=22.5$ mag, using the $ugri$ catalog. The bright magnitude cut is
adopted to reduce the scatter due to increased photometric errors at
fainter magnitudes. The diagrams show that the sequence of
GCs/UCDs/stars in the \gi-\gr matches with the sequence of extended
objects, while in the \gi-\ur diagram the degeneracy is less dramatic,
making more efficient the separation of compact/extended sources.

 To obtain a rough estimate of the increase of contamination due to
 the lack of $u$-band photometry we proceed as follows. Using the
 $ugri$ catalog in the FDS area, we adopt the GC selection scheme
 described in the previous section but use the \gi-\gr color
 combination, instead of the \gi-\ur, for the selections on the
 color-color plane. y comparing the number of GC candidates
 identified using the \gi-\ur color-color, $N_{ugri}^{GC}$, with the
 number of candidates identified using \gi-\gr color-color,
 $N_{gri}^{GC}$, we find
 $(N_{gri}^{GC}-N_{ugri}^{GC})/N_{gri}^{GC}\sim0.48$. Therefore, this
 single change in the criteria for GC selections implies the number of
 sources identified as GC candidates is nearly doubled over the FDS
 area. Such increase is not spatially uniform: it is close to
 $\sim80\%$ in background regions, i.e. far from bright galaxies and
 their host GC system, and drops to $\sim15\%$ around bright galaxies.
 This difference shows that GC selection in the central cluster area,
 where GCs have a high surface density, is already quite efficient
 with a 3-band combination. In contrast, the addition of the $u$-band
 makes a significant difference in GC selection in the outer parts of
 the cluster where the fractional background contamination, mostly due
 to MW stars, is higher.

In spite of the higher level of contamination, the FDSex $gri$-band
catalog also includes the area of NGC\,1316, Fornax A, the brightest
galaxy in the cluster in optical bands, a peculiar giant elliptical,
suggested being in its second stage of mass assembly
\citep{iodice17b}. It is then of particular interest to show here, for
the first time, the global properties of the GCs over such wide area.
We should, however, be aware that NGC\,1316 is known to contain
relatively young GCs \citep[$\sim2-3$ Gyr,
  e.g.][]{gomez01,goudfrooij01b,sesto17}, which are not part of our
reference sample. Young GCs are in general bluer and brighter than
equally massive old GCs; hence, we bear in mind that our selection is
intrinsically biased toward old GCs.

\begin{sidewaysfigure*}
   \centering
   \includegraphics[trim={7.5cm 3cm 7cm 3cm},clip,scale=0.085]{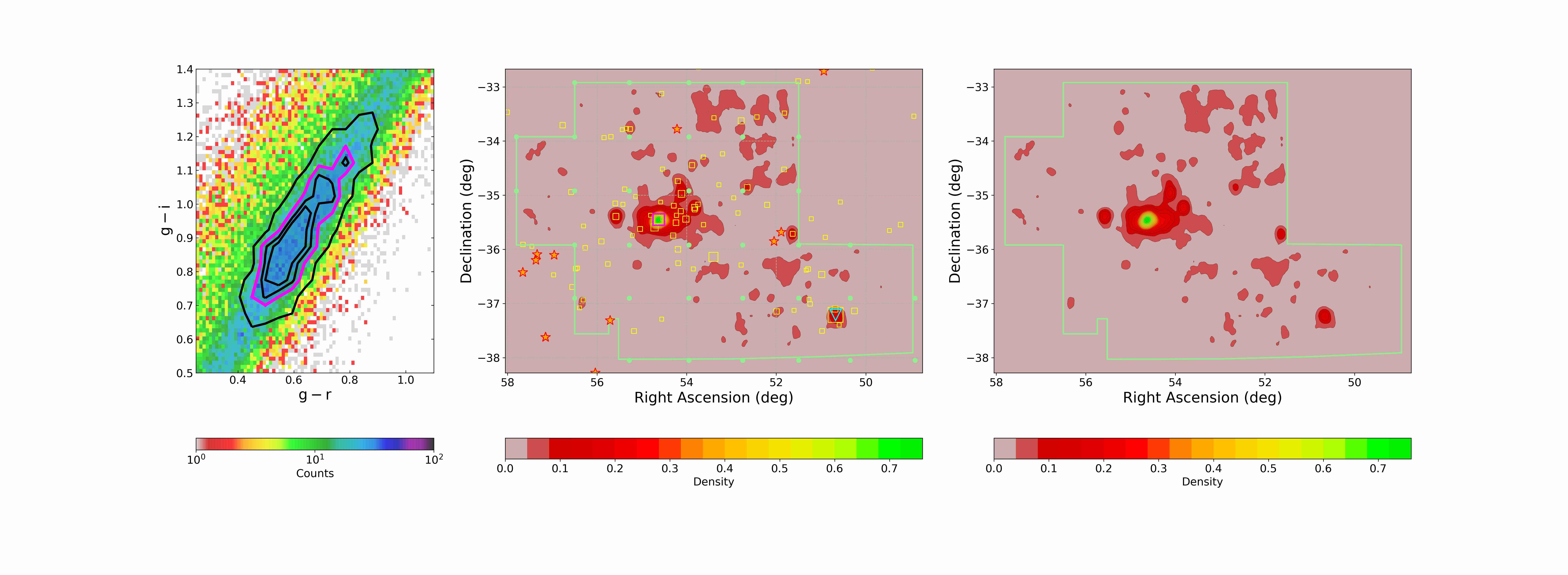}
      \caption{Same as in Figure \ref{mappa_gc}, except that over the
        FDSex area, and the \gi-\gr color-color diagram is used for
        GCs selection. The position of NGC\,1316 is shown with
        light-blue empty triangle in the middle panel.}
         \label{mappa_gc_gri}
   \end{sidewaysfigure*}

Using the same procedures described in the previous sections, except
that \gr is used instead of \ur, we analyze the surface distribution
maps over the 27 sq. degrees of the FDSex area. For sake of clarity in
Figure \ref{mappa_gc_gri} we only show the second color-color density
contour, corresponding to the iso-density contour level of 15 GCs from
the master catalog.  In the panels of the figure, no obvious GC
substructure appears bridging the core of the main Fornax cluster to
the Fornax A sub-group. The two brightest galaxies, NGC\,1399 and
NGC\,1316, are $\sim3.6\deg$ apart ($\sim1.3$ Mpc) and the density map
of the $\sim10.200$ GCs selected does not reveal any hint of residual
GC tails along the direction connecting the bright ellipticals, with
the possible only exception of the East-West elongation of GCs around
the cluster core still visible in the $gri$ map, although with less
details compared with the $ugri$ maps.

The higher level of contamination of the $gri$ maps appears in some
spurious features. Figure \ref{mappa_gc_gri} shows a structure around
the area of coordinates [R.A.=52 deg, Dec.=-34 deg], characterized by
nearly the same geometric appearance of the FDS fields \#14, \#19 and
\#31.  Such structure is completely unseen in the $ugri$ maps which
also cover the area; inspecting the three FDS fields we find slightly
deeper limiting magnitudes and slightly poorer source compactness
relative to the neighboring fields: combined, two effects
generate larger number of detections with poorer morphologic
characterization, hence a higher fraction of GCs contamination.

To have a less contaminated sample, we narrowed the sample of GC
candidates by using a brighter magnitude cut, more stringent ranges on
the various morphological parameters in Table \ref{tab_params}, and
narrower color-color regions. Using narrower selections, the spurious
structure around the fields FDS\#14/19/31 disappears. Nevertheless, no
matter how much the GC sample is narrowed with more strict selections,
no GC substructure emerges along the NGC\,1316/NGC\,1399 direction.

\begin{figure}[ht]
   \centering
   \includegraphics[trim={1.5cm 0cm 2cm 0.5cm},clip,scale=0.4]{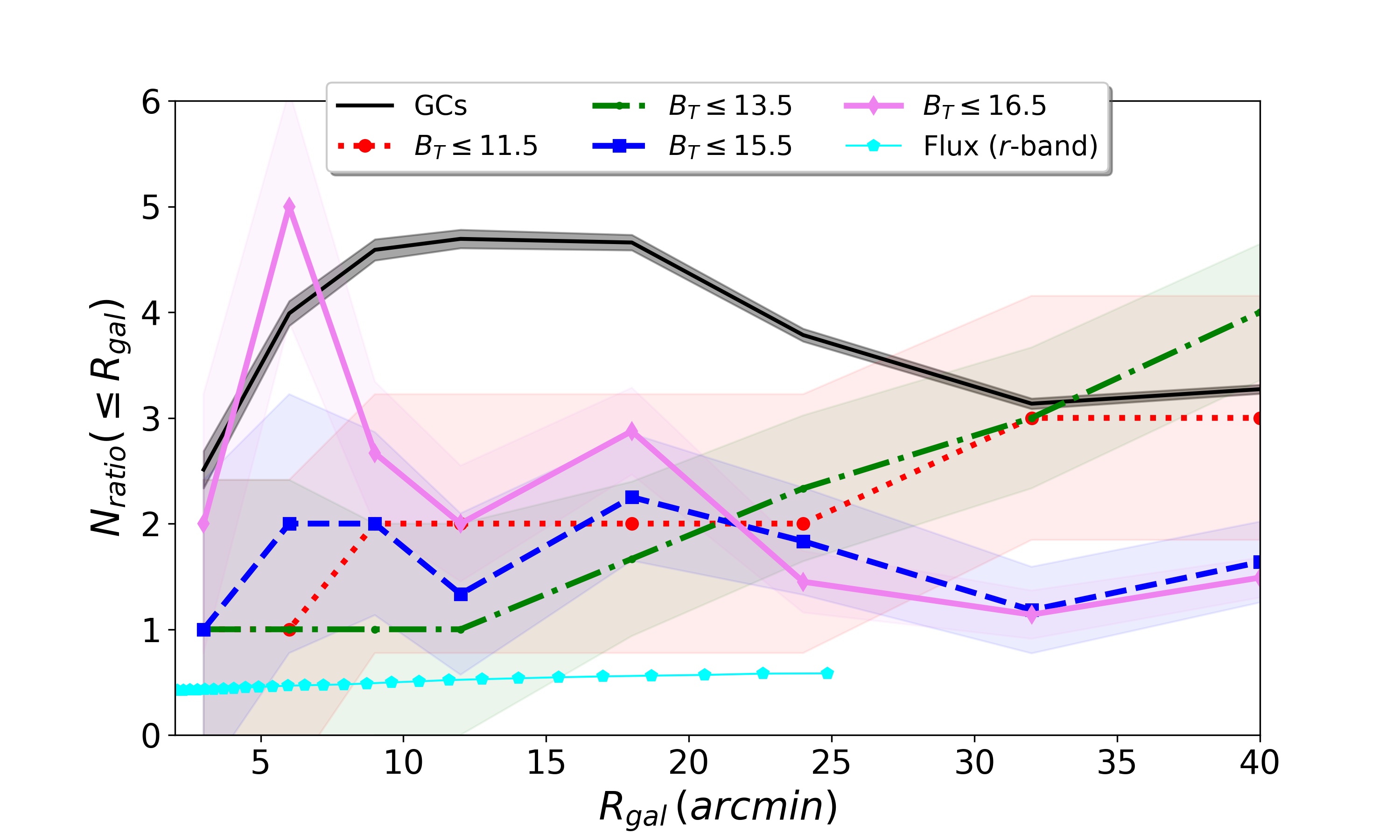}
      \caption{Number ratio of the total number of sources around
        NGC\,1399 and NGC\,1316 within a given galactocentric radius
        in the respective environment: $N_{ratio}(\leq
        R_{gal})=N_{Sources}^{N1399}/N_{Sources}^{N1316}$.  The number
        ratio for GCs is shown with a black solid line; number ratios
        for galaxies at a given bright magnitude cut are also shown,
        and labeled. The $r$-band flux ratio between the two galaxies
        within $R_{gal}$ is shown with light-blue solid line and
        pentagons.}
         \label{nratio}
   \end{figure}

By counting the number of GCs candidates within a given radius
centered on each of the two bright galaxies within the respective
environments, we find that the number of GCs around NGC\,1399
outnumbers NGC\,1316 by a factor of 4-4.7 at galactocentric radii
$R_{gal}$ of $\sim6\arcmin$ and $\sim24\arcmin$, and by a factor of
$\sim3$ out to $R_{gal}\sim40\arcmin$.  Figure \ref{nratio} shows the
number ratio $N_{ratio}^{Sources}(\leq
R_{gal})=N^{Sources}_{N1399}/N^{Sources}_{N1316}$ versus
galactocentric distance for GCs candidates (black line in the figure),
for galaxies brighter than a given limit (as labeled in the figure),
and the flux ratio of the $r$-band integrated magnitudes of the two
galaxies \citep[from][light-blue line in the figure]{iodice16,
  iodice17b}.

The median $N_{ratio}^{galaxies}$ for galaxies in the range of
$6\arcmin -24\arcmin$ is $\sim2.0$ with $rms_{MAD}=0.3$.  Assuming a
nearly uniform contamination of the FDSex catalogs around the two
regions, we estimate the overdensity of GCs around NGC\,1399 compared
to NGC\,1316 ($N_{ratio}^{GCs}\sim4$) is a factor of $\sim2$ larger
than the overdensity of galaxies in the magnitude range $11.5\leq
B_T~(mag)\leq 16.5$ ($-20 \leq M_{B,tot}~(mag) \leq -15$). Hence, even
accounting for the larger density of bright and faint galaxies of all
morphological types, the population of GCs is considerably larger in
the region of $6\arcmin \leq R_{gal} \leq 24\arcmin$ around NGC\,1399
compared with NGC\,1316, and mainly composed of blue GCs.

This overpopulation of GCs is likely associated with the intra-cluster
GCs component; on the contrary, the relative GCs under-density around
Fornax A, and the lack of any major accretion events of NGC\,1316
  that could have significantly increased the specific frequency of
  blue GCs, is possibly at the basis of the lack of any significant
GC substructure. Furthermore, as expected from the known factor of
$\sim2$ higher total magnitude of NGC\,1316 compared to NGC\,1399, the
$r$-band flux ratio between the two ellipticals is
$\sim0.4\pm0.1$ (light--blue line in Figure \ref{nratio}), a factor of
$8-10$ lower than the GCs count ratio.

Figure \ref{nratio} also shows some other features : $a)$ the
GCs and bright galaxies with $B_T\leq11.5$ mag and $B_T\leq13.5$ mag
($M_{B,tot}=-20$ and $-18~mag$, respectively) have
$N_{ratio}^{galaxy}\sim3$ at $R_{gal}\geq 30\arcmin$, while for the
fainter galaxy bin limits we find $N_{ratio}^{galaxy}\sim1.3$; $b)$
the nearly flat GCs $N_{ratio}$ within $9\leq R_{gal}~(\arcmin)\leq
20$, which assumes a value of $4.66\pm0.04$.  A more detailed analysis
of such properties combined with the data in other galaxy
  clusters is in progress (Cantiello et al., 2020, in prep.)

\section{FDS catalogs of background sources and related science}\label{ctortora}

The depth and spatial resolution of the FDS images, together with
ancillary data from other spectral ranges available in this field,
provide the opportunity to study the stellar populations and
structural properties of galaxies beyond the cluster, as well as to
discover rare astrophysical objects, like compact massive galaxies and
strong gravitational lenses (e.g., \citealt{Tortora+18_UCMGs};
\citealt{Petrillo+17_CNN}). The FDS image quality is similar to the
one of the KiDS survey \citep{kuijken19}, since the longer exposures
in FDS images are balanced by a slightly poorer seeing ($r$-band FWHM
of $\sim 0\farcs9$ for FDS, vs $\sim 0\farcs65$ in KiDS). The limiting
magnitudes in the two surveys are quite similar, but FDS is deeper in
$i$-band.

 Taking advantage of the FDS data, we aim at determining photometric
 redshifts, stellar masses, galaxy classifications and structural
 parameters of thousands of background galaxies. This will provide a
 complete characterization of the background galaxy population over
 the area, for investigating the evolution of the structural and
 stellar properties of galaxies as a function of redshift and
 mass. The tools for deriving all required quantities are already
 available and well tested in our team
 \citep[e.g.][]{labarbera08,Cavuoti+17_METAPHOR,roy18}.

\begin{figure}
\centering
\includegraphics[width=0.2\textwidth]{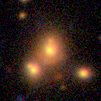}
\includegraphics[width=0.2\textwidth]{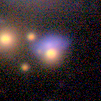}
\caption{Two example lens candidates found in the FDS fields applying
  the CNN code; the image cutout have $20\farcs0$ side. Left:
  FDSJ032720.32-365821.81, right: FDSJ034739.60-352516.23
}\label{fig:lenses}
\end{figure}

As a first test on the background galaxy population in FDS, we run a
code to find galaxy-galaxy lens candidates. The code uses a machine
learning classification method based on Convolutional Neural Networks
(CNNs), and was already applied to the KiDS survey
(\citealt{deJong+17_KiDS_DR3,Petrillo+19_CNN}). We performed the run
of CNN with a network trained on a large sample of $r$-band (or
combined $g$-$r$-$i$) KiDS images, to the equivalent FDS
images. Although the network is not customized and trained on FDS
images, KiDS and FDS are based on data from the same telescope and
camera, and, as mentioned above, are comparable in both FWHM and
depth. Therefore, this is a valid approach to search gravitational
lenses in FDS.  In fact we already discovered several
  gravitational lens candidates in the FDS fields, two of them, with
FDS catalogue ID and coordinates: FDSJ032720.32-365821.81 at
[51.834682; -36.972725] and FDSJ034739.60-352516.23 at [56.91502,
  -35.421176], are presented in \Fig\ref{fig:lenses} as an example of
the potential of this approach.

\section{Conclusions}
\label{summary}

We have presented the photometric and morphometric catalog of compact
and slightly extended sources in the Fornax galaxy cluster,
derived with VST observations within the FDS survey over an area of
$\sim21$ square degrees in $ugri$-bands, and in $gri$-bands for a
total of $\sim27$ square degrees.

The $ugri$ data of FDS cover the main body of Fornax, centered on
NGC\,1399, and extend out to $\sim1~Mpc$, the virial radius of the
cluster. The $gri$ coverage, FDSex, extends to the South-West region
of the Fornax A sub-cluster with its brightest galaxy, NGC\,1316.

Because of the large FWHM variation from field to field, to improve
the uniformity of sources detection and their morphological
characterization, we derived a master-detection frame by coadding all
$gri$ single exposures with $FWHM\leq0\farcs9$; starting from a median
FWHM ranging from $0\farcs92$ to $1\farcs26$ with $rms$ within
$0\farcs11-0\farcs17$ for the various bands, adopting the multi-band
stacking procedure we ended up with a master-detection frame with a
median FWHM of $0\farcs80\pm0\farcs04$, $\sim15\%$ improvement over
the median FWHM of the highest resolution imaging ($r$-band), and a
factor of $\sim2.5$ lower $rms$.

We calibrated the photometry using a two-step procedure, to reduce the
effect of the independent calibration of the FDS fields, which
generate a non-negligible photometric offset between neighboring
fields. The first calibration step follows the standard calibration
plan of VST frames. As a second step, we used the APASS photometry to
derive a matrix to match the full FDS catalog to a unique
reference. With this approach, the photometric offset between fields
becomes negligible, and the re-calibrated photometry shows a general
good match to existing literature data from SKyMapper, from the
HST/ACSFCS survey and to predictions from stellar population synthesis
models.

The catalogs are available through the project web pages, and will
also be available on CDS. In the catalogs we provide the position, the
photometry and the morphometry for 1.7 million sources with $ugri$
detections, and for 3.1 million sources with $gri$ data.

As a preliminary use of the catalogs, we analysed the 2-D distribution
of compact stellar systems in the area, with particular care to GCs.

With the FDS instrumental setup, and at the distance of Fornax, GCs
are by all means point-like sources, except for a possible fraction of
$\lsim0.5\%$ of the population. Hence, GCs can be identified by their
compactness. 

To obtain the least contaminated GCs sample, we selected a number of
morpho-photometric features, and analysed them over a reference
catalog of confirmed GCs and UCDs in Fornax. Such catalog is build by
cross-matching the FDS catalogs with available spectroscopic and
photometric datasets of confirmed GC/UCD. The reference catalog is
then used to define the GCs loci in the parameter space, for the
chosen photometric and morphometric parameters.

The GCs maps over the FDS area confirm the results of previous
studies, about the presence of a large inter-galactic GC population
around the main body of the cluster, centered on NGC\,1399, stretched
along the East-West direction. Here we find a small tilt of the
distribution in the direction of  NGC\,1336 by $\sim10\deg$. The
distribution appears to extend over $\sim1~Mpc$ from side to side,
highly flattened, with an ellipticity of $\sim0.65$. In addition to
our previous results, we find that one of the features already
discussed, which extends from the main cluster body to the South-West
direction, might be a tail of relatively blue GCs from NGC\,1404, a
bright galaxy close to the cluster core and with a peculiarly poor GCs
population.

Of the GCs features already commented in the past, we here highlight
the case of NGC\,1336, which we confirm to be relatively
isolated from the cluster, and with a high specific frequency of GCs;
this might support the hypothesis that it is an infalling massive
galaxy, with a GCs system that possibly experienced only few
disruption processes. We also inspected the blue/red GCs maps, and
confirm the known property of blue GCs residing in the wider cluster
area, and red GCs being more concentrated on massive galaxies.

Systems selected to fit the color-magnitude range of spectroscopically
confirmed UCDs show a substantial overdensity in the central
cluster. The 160 UCD candidates are about three times more than the
currently known UCDs in Fornax and would require spectroscopic follow
up to learn more about their nature.

We also derived the GCs maps over the FDSex area, which has the
disadvantage of suffering for larger contamination because of the lack
of $u$-band over the NGC\,1316 area, but has the advantage of covering
this brightest cluster galaxy. With the caveat that the $gri$ catalogs
do not allow the detailed analysis allowed over the FDS area, despite
our attempts to obtain a cleaner GC candidates sample, we do not find
significative GCs structures along the NGC\,1399-NGC\,1316 direction,
which extends over a projected distance of $\sim1.3~Mpc$. This might
be due to the lower efficiency of the GC identification. However,
assuming similar contamination of the $gri$ catalogs over the
NGC\,1399 and NGC\,1316, we find that the GC population of the former
outnumbers the second by a factor of $\sim 4$, and by about a factor
of $\sim10$ when normalized to galaxy luminosity, within a
galactocentric range of $6\arcmin-24\arcmin$, and remains a factor of
$\sim3$ higher than NGC\,1316 at larger galactocentric radii, out to
$\sim40\arcmin$. Hence, the 'contrast' of the GC populations towards
NGC\,1316 might be too low for the purpose of our study, -- in spite
of its luminosity twice larger than NGC\,1399 -- and might explain the
difficulty in finding GCs sub-structures, which intrinsically need a
large number of candidates over a given region to be identified. The
rich intra-cluster GCs population around NGC\,1399 does not seem to be
matched by a similarly rich system around NGC\,1316, the brightest
galaxy of the Fornax A sub-cluster.  In spite of this, the lack of
obvious GC sub-structures between these two bright and massive
galaxies might also be consequence of the NGC\,1316 sub-cluster being
in its first infalling phase and evolving autonomously, a result also
supported by an independent analysis of FDS data for galaxy surface
brightness profiles and intracluster light \citep{iodice17b,raj19}.

A deeper analysis of the 2-D maps, and other characteristics of the
GCs over the FDS and FDSex area is in progress, and will be presented
in a dedicated paper.

We finally presented an example use of the catalogs to analyze
background galaxies. Using machine learning methods, already tested on
the KiDS survey with VST, we identified two lens candidates in the FDS
area.

\begin{acknowledgements}
This research was made possible through the use of the AAVSO
Photometric All-Sky Survey (APASS), funded by the Robert Martin Ayers
Sciences Fund and NSF AST-1412587.

This work is based on visitor mode observations collected at the
European Organisation for Astronomical Research in the Southern
Hemisphere under the following VST GTO programs: 094.B-0512(B),
094.B-0496(A), 096.B-0501(B), 096.B-0582(A).

INAF authors acknowledge financial support for the VST project
(P.I. P. Schipani).  We acknowledge the use of data from the SKyMapper
survey. The national facility capability for SkyMapper has been funded
through ARC LIEF grant LE130100104 from the Australian Research
Council, awarded to the University of Sydney, the Australian National
University, Swinburne University of Technology, the University of
Queensland, the University of Western Australia, the University of
Melbourne, Curtin University of Technology, Monash University and the
Australian Astronomical Observatory. SkyMapper is owned and operated
by The Australian National University's Research School of Astronomy
and Astrophysics. The survey data were processed and provided by the
SkyMapper Team at ANU. The SkyMapper node of the All-Sky Virtual
Observatory (ASVO) is hosted at the National Computational
Infrastructure (NCI). Development and support the SkyMapper node of
the ASVO has been funded in part by Astronomy Australia Limited (AAL)
and the Australian Government through the Commonwealth's Education
Investment Fund (EIF) and National Collaborative Research
Infrastructure Strategy (NCRIS), particularly the National eResearch
Collaboration Tools and Resources (NeCTAR) and the Australian National
Data Service Projects (ANDS).

MP acknowledges financial contribution from the agreement ASI-INAF
n.2017-14-H.O. JFB acknowledges support through the RAVET project by
the grant AYA2016-77237-C3-1- P from the Spanish Ministry of Science,
Innovation and Universities (MCIU) and through the IAC project TRACES
which is partially supported through the state budget and the regional
budget of the Consejer\'ia de Econom\'ia, Industria, Comercio y
Conocimiento of the Canary Islands Autonomous Community.  GvdV
acknowledges funding from the European Research Council (ERC) under
the European Union's Horizon 2020 research and innovation programme
under grant agreement No 724857 (Consolidator Grant ArcheoDyn). CT
acknowledges funding from the INAF PRIN-SKA 2017 program 1.05.01.88.04.
\end{acknowledgements}



\bibliographystyle{aa}
\bibliography{cantiello_mar20,myrefs_CT} 

\end{document}